\documentclass[conference]{IEEEtranICDE}
\usepackage{balance}
\usepackage{setspace}
\usepackage{multirow}
\usepackage{hhline}

\usepackage{wrapfig}
\usepackage{amssymb}
\usepackage{amsthm}
\let\labelindent\relax
\usepackage{enumitem}
\usepackage{listings}
\usepackage{epstopdf}
 \usepackage[noadjust]{cite}

\usepackage[utf8]{inputenc}
\usepackage[english]{babel}

\usepackage[usenames, dvipsnames]{color}

\usepackage{amssymb}

\usepackage{xspace}

\usepackage{fancyhdr}
\usepackage{csquotes}
\usepackage[boxed,ruled,vlined,linesnumbered]{algorithm2e}
\usepackage{url}
\usepackage{framed}
\usepackage{caption}

\usepackage{subcaption}

\usepackage{times}
\usepackage{graphicx}
\newcommand{\B}{\vspace*{-\smallskipamount}}
\newcommand{\BB}{\vspace*{-\medskipamount}}
\newcommand{\BBB}{\vspace*{-\bigskipamount}}

\SetAlgoSkip{}

\IEEEoverridecommandlockouts

\usepackage{times,amsmath,epsfig,amsthm}

\title{Partitioned Data Security on Outsourced Sensitive and Non-sensitive Data}

\author{\IEEEauthorblockN{Sharad Mehrotra$^1$, Shantanu Sharma$^1$, Jeffrey D. Ullman$^2$, and Anurag Mishra$^1$\thanks{\textbf{Accepted in IEEE International Conference on Data Engineering (ICDE), 2019.} For the final version, please refer to the conference proceeding.  \newline  This material is based on research sponsored by DARPA under agreement number FA8750-16-2-0021. The U.S. Government is authorized to reproduce and distribute reprints for Governmental purposes notwithstanding any copyright notation thereon. The views and conclusions contained herein are those of the authors and should not be interpreted as necessarily representing the official policies or endorsements, either expressed or implied, of DARPA or the U.S. Government. This work is partially supported by NSF grants 1527536 and 1545071. \newline The authors are thankful to the PC chairs, vicechairs, and reviewers for helping in improving the presentation of the paper.}}
\IEEEauthorblockA{$^1$University of California, Irvine. $^2$Stanford University, USA.}
}

\begin{document}
\maketitle
\begin{abstract}
Despite extensive research on cryptography, secure and efficient query processing over outsourced data remains an open challenge. This paper continues along the emerging trend in secure data processing that recognizes that the entire dataset may not be sensitive, and hence, non-sensitivity of data can be exploited to overcome limitations of existing encryption-based approaches. We propose a new secure approach, entitled query binning (QB) that allows non-sensitive parts of the data to be outsourced in clear-text while guaranteeing that no information is leaked by the joint processing of non-sensitive data (in clear-text) and sensitive data (in encrypted form). QB maps a query to a set of queries over the sensitive and non-sensitive data in a way that no leakage will occur due to the joint processing over sensitive and non-sensitive data. Interestingly, in addition to improve performance, we show that QB actually strengthens the security of the underlying cryptographic technique by preventing size, frequency-count, and workload-skew attacks.
\end{abstract}

\section{Introduction}
\label{sec:introduction}
The last two decades have witnessed the development of secure and privacy-preserving encryption-based~\cite{gentry2009fully,DBLP:conf/sp/SongWP00,DBLP:conf/stoc/Goldreich87} 
or secret-sharing-based~\cite{DBLP:journals/cacm/Shamir79,DBLP:journals/isci/EmekciMAA14,DBLP:conf/eurocrypt/GilboaI14,DBLP:conf/eurocrypt/BoyleGI15} 
techniques to realize the database as a service model.
Despite significant progress, a cryptographic approach that is both \emph{secure} (\textit{i}.\textit{e}., no leakage of sensitive data to the adversary) and \emph{efficient} (in terms of time) simultaneously has proved to be very challenging. Broadly, work on cryptography to support secure outsourcing has taken the following directions:

\parskip 0pt
\setlength{\parindent}{15pt}

\noindent\emph{\textbf{Techniques that support strong security guarantees}}. The leading example of which is fully homomorphic encryption~\cite{gentry2009fully}, which when mixed with oblivious-RAM (ORAM)~\cite{DBLP:conf/stoc/Goldreich87}, offers possibly amongst the most secure mechanisms. However, such mechanisms incur high computational overhead. 

\noindent\textit{\textbf{Techniques that do not depend on the data encryption}} but provide strong security, especially, information-theoretic security, by distributing a value in the form of the secret-shares to non-colluding clouds. Shamir's secret-sharing~\cite{DBLP:journals/cacm/Shamir79}, distributed point functions (DPF)~\cite{DBLP:conf/eurocrypt/GilboaI14}, and function secret-sharing (FSS)~\cite{DBLP:conf/eurocrypt/BoyleGI15} are a few examples of such techniques. Such methods often limit the type of operations one can perform while imposing high overhead in terms of communication.

  \noindent\textit{\textbf{Techniques that try to support a wide range of operations}} including index-based retrieval or joins, such as CryptDB~\cite{DBLP:journals/cacm/PopaRZB12}, Arx~\cite{arx-popa-2017}, searchable encryption~\cite{DBLP:conf/sp/SongWP00}, and cryptographic indexes~\cite{DBLP:conf/dbsec/ShmueliWEG05}. 
   Such techniques often trade security for performance; for instance, techniques that depend on deterministic and order-preserving encryptions, traversal of the index by the cloud, or leakage of the searching token do not offer strong security. Papers~\cite{DBLP:conf/ccs/NaveedKW15,DBLP:conf/ccs/KellarisKNO16} show that order-preserving and deterministic encryption techniques when used together, on a dataset in which the entropy of the values is not high enough can leak the entire data in clear-text to an attacker through frequency analysis on the encrypted data.

  \noindent\textit{\textbf{Techniques/systems that exploit secure hardware}} (Intel Software Guard Extensions (SGX)~\cite{costan2016intel}), \textit{e}.\textit{g}., M2R~\cite{DBLP:conf/uss/DinhSCOZ15}, VC3~\cite{DBLP:conf/sp/SchusterCFGPMR15}, Opaque~\cite{opaque}, and EnclaveDB~\cite{DBLP:conf/sp/PriebeVC18}. Such techniques also leak information during a query execution due to different attacks on SGX (\textit{e}.\textit{g}., cache-line, branch shadowing, and page-fault attacks~\cite{DBLP:conf/ccs/WangCPZWBTG17,DBLP:conf/eurosec/GotzfriedESM17}) and are significantly slow when overcoming these attacks using ORAM-based computations or emerging architectures such as T-SGX~\cite{DBLP:conf/ndss/Shih0KP17} or Sanctum~\cite{DBLP:conf/uss/CostanLD16}.

\parskip 0pt
\setlength{\parindent}{15pt}


Given the state of the research, this paper explores a radically different approach to secure outsourcing that scales cryptographic mechanisms using database techniques while providing strong security guarantees. Our work is motivated by recent works on the hybrid cloud that has exploited the fact that for a large class of application contexts, data can be partitioned into sensitive and non-sensitive components~\cite{DBLP:conf/ccs/ZhangZCWR11,DBLP:conf/sigmod/OktayMKK15,TR}. Such a classification, which is common in industries for secure computing~\cite{url2,url3} and done via appropriately using existing techniques surveyed in~\cite{DBLP:journals/sigkdd/FarkasJ02}; for example, (\textit{i}) inference detection using graph-based semantic data modeling~\cite{DBLP:conf/sp/Hinke88}, (\textit{ii}) user-defined relationships between sensitive and non-sensitive data~\cite{smith1990modeling}, (\textit{iii}) constraints-based mechanisms, (\textit{iv}) sensitive patterns hiding using sanitization matrix~\cite{lee2004hiding}, and (\textit{v}) common knowledge-based association rules~\cite{DBLP:conf/dasfaa/LiSY07}. However, it is important to mention here that non-sensitive data can, over time, become sensitive and/or lead to inferences about sensitive data. This is an inevitable risk of the approaches that exploit sensitive data classification. Note that all the above-mentioned work based on sensitive/non-sensitive classification make a similar assumption. Indeed, another way to view this assumption is that today, cloud solutions, already outsource databases without encryption and are risking the loss of not just non-sensitive data but also sensitive data.

Based on data classification into sensitive and non-sensitive data, secure solutions for hybrid cloud have been developed~\cite{DBLP:conf/ccs/ZhangZCWR11,DBLP:conf/sigmod/OktayMKK15,TR}. These solutions outsource only non-sensitive data and enjoy both the benefits of the public cloud as well as strong security guarantees (without revealing sensitive data to an adversary). While these techniques provide an effective and secure solution, they are, however, requiring data owners to maintain potentially unbounded storage locally and also suffer from significant inter-cloud communication overheads.

Our goal, in this paper, is to explore how sensitive and non-sensitive classification can be exploited by secure data processing techniques that store the entire data in the public cloud to bring new efficiencies to secure data processing. In particular, in the envisioned public cloud model, data is stored in a partitioned way -- sensitive data is secured using any existing cryptographic technique (unlike the hybrid cloud solution where the owner stores the sensitive data) and non-sensitive data resides in plaintext. Query processing is also split into encrypted and plaintext query processing. We refer to this as \emph{partitioned computing}. Unlike the case of the hybrid cloud, when implementing partitioned computing at a public cloud, data processing performed on the sensitive and non-sensitive parts of the data may reveal exact encrypted tuples 
and cleartext tuples that satisfy the query to the adversary. Consequently, this leads to inferences about sensitive data, which will be explained in detail in \S\ref{sec:Preliminaries_and_Problem_Statements}.

We define a security model (\S\ref{subsec:Security Definition and Correctness}) that formally states what it means to be secure in partitioned computing. We then develop a query binning (QB) approach that realizes secure partitioned computing for selection queries. We focus on selection queries for several reasons. First, selection queries are important in their own right. For instance, several key-value stores (\textit{e}.\textit{g}., Amazon Dynamo) and document stores (\textit{e}.\textit{g}., MongoDB) focus exclusively on selection queries (with limited support for joins). Furthermore, most cryptographic research has also focused on selection queries~\cite{gentry2009fully,DBLP:conf/sp/SongWP00,DBLP:conf/stoc/Goldreich87,DBLP:conf/eurocrypt/GilboaI14}. Since our goal is to speed up existing cryptographic techniques (and not to extend their functionality and make them resilient against attacks, such as order-revealing, inferences from deterministic encryptions, leakages from SGX, and different side-channel attacks~\cite{DBLP:conf/ccs/NaveedKW15,DBLP:conf/ccs/KellarisKNO16,DBLP:conf/eurosec/GotzfriedESM17}), 
 we focus on selection queries. Nonetheless, there are recent work on cryptographic joins~\cite{DBLP:journals/tods/PangD14} and also on joins using SGX~\cite{opaque}. These approaches, however, are not yet practical, \textit{e}.\textit{g}., from the efficiency perspective, Opaque~\cite{opaque} takes 89 seconds to execute a selection query on a dataset of size 700MB. The same query takes about 0.2 milliseconds over cleartext processing. Also, systems, \textit{e}.\textit{g}., Opaque, support limited operations (only primary-to-foreign key joins) and, furthermore, leaks information due to cache-line, page table-based, branch shadowing, and output-size attacks~\cite{DBLP:conf/ccs/WangCPZWBTG17,DBLP:conf/eurosec/GotzfriedESM17}. Many of these attacks can be overcome with expensive ORAM techniques, and the QB approach alongside such approaches can be exploited to improve efficiency.


We show two interesting effects of using QB: (\textit{i}) By avoiding cryptographic processing on non-sensitive data, \emph{the joint cost of communication and computation of QB is significantly less than the computation cost of a strongly secure cryptographic technique}\footnote{QB trades off increased communication costs for executing queries, while reducing very significantly cryptographic operations. This tradeoff significantly improves performance, especially, when using cryptographic mechanisms, \textit{e}.\textit{g}., fully homomorphic encryption that takes several seconds to compute a single operation~\cite{DBLP:journals/csur/MartinsSM17}, secret-sharing-based techniques that take a few seconds~\cite{DBLP:journals/isci/EmekciMAA14},
or techniques such as bilinear maps that take over 1.5 hours to perform joins on a dataset of size less than 10MB~\cite{DBLP:journals/tods/PangD14}. When considering such cryptography, increased communication overheads are fully compensated by the savings. A similar observation, albeit in a very different context was also observed in~\cite{DBLP:conf/sigmod/OktayMKK15} in the context of MapReduce, where overshuffling to prevent the adversary to infer sensitive keys in the context of hybrid cloud was shown to be significantly better compared to private side operations.} (\textit{e}.\textit{g}., homomorphic encryptions, DPF~\cite{DBLP:conf/eurocrypt/GilboaI14}, or secret-sharing-based technique~\cite{DBLP:journals/isci/EmekciMAA14}) 
on the entire encrypted data; and hence, QB improves the performance of strong cryptographic techniques over a large-scale dataset (\S\ref{sec:Experiments}). (\textit{ii}) 
\emph{QB provides an enhanced security by preventing several attacks such as output size, frequency-count, and workload-skew attacks, even when the underlying cryptographic technique is susceptible to such attacks} (\S\ref{sec:Desiderata}).

\smallskip
\noindent\textbf{Contributions.} The primary contributions of this paper are: (\textit{i}) A formal definition of \emph{partitioned data security} when jointly processing sensitive and non-sensitive data (\S\ref{subsec:Security Definition and Correctness}). (\textit{ii}) An efficient QB approach (\S\ref{sec:Query Bucketization}) that guarantees partitioned data security, supporting cloud-side-indexes, and that can be built on top of any cryptographic technique. (\textit{iii}) An analytical model and experimental validation to show the effectiveness of QB over a strong secure cryptographic technique (\S\ref{sec:Experiments}). (\textit{iv}) A weak cryptographic technique (\textit{e}.\textit{g}., cloud-side indexable techniques~\cite{DBLP:conf/dbsec/ShmueliWEG05,arx-popa-2017} 
) becomes secure and efficient when mixed with QB (\S\ref{sec:Desiderata}).

\smallskip
\noindent\textbf{Full version.}~\cite{TR2018} provides the full version of this paper. The full version provides: (\textit{i}) formal security and computational complexity proofs of QB, (\textit{ii}) extensions of QB to deal with non-identical searchable attribute-based column-level sensitivity, join, and range queries, (\textit{iii}) some additional experiments to show insert and the use of indexable cryptographic techniques, and (\textit{iv}) an analytical formal security model to compare QB with a pure cryptographic technique under different conditions and different security levels such as preventing size, frequency-count, and workload-skew attacks. QB can also be extended to support group-by aggregation queries as well; however, extending it to support nested queries is more complex and will need a significant extension.

\smallskip
\noindent\textbf{Related work on secure selection queries.} Broadly, existing research on secure selection query execution techniques can be classified into four categories, as follows: (\textit{i}) \emph{Encryption-based techniques} examples of which include order-preserving encryption,
deterministic encryption, homomorphic encryption~\cite{gentry2009fully}, searchable encryption~\cite{DBLP:conf/sp/SongWP00}, and
ORAM~\cite{DBLP:conf/stoc/Goldreich87}. 
(\textit{ii}) 
\emph{Secret-sharing~\cite{DBLP:journals/cacm/Shamir79} based techniques} that include DPF~\cite{DBLP:conf/eurocrypt/GilboaI14}, FSS~\cite{DBLP:conf/eurocrypt/BoyleGI15}, and~\cite{DBLP:journals/isci/EmekciMAA14}. (\textit{iii}) 
\emph{Trusted-hardware-based techniques} that include
~\cite{DBLP:conf/sp/SchusterCFGPMR15,DBLP:conf/uss/DinhSCOZ15,opaque,DBLP:conf/sp/PriebeVC18}. (\textit{iv}) 
\emph{Sensitivity-based techniques.} MapReduce~\cite{DBLP:conf/ccs/ZhangZCWR11,DBLP:conf/sigmod/OktayMKK15} and SQL data processing~\cite{TR}. Both MapReduce and SQL execution solutions work on the principle of sensitivity-based data partitioning over the hybrid cloud.

\parskip 0pt
\setlength{\parindent}{15pt}

Each of the above strategies has resulted in corresponding systems that support secure data processing; \textit{e}.\textit{g}.,
CryptDB~\cite{DBLP:journals/cacm/PopaRZB12}, Arx~\cite{arx-popa-2017}, and Opaque~\cite{opaque} are some novel encryption-based systems. Likewise, Microsoft Always Encrypted, Oracle 12c, Amazon Aurora, and MariaDB are industrial secure encrypted databases. DSSE-based SDB~\cite{pulsar} is a secret-sharing and encryption-based system while Arx~\cite{arx-popa-2017} and Opaque~\cite{opaque} work on the data sensitivity principle.


These systems/techniques are unable to prevent one or more of the following attacks: (\textit{i}) size attack, \textit{i}.\textit{e}., an adversary having some background knowledge can deduce the full/partial outputs by simply observing the output sizes~\cite{opaque}; (\textit{ii}) frequency attack, \textit{i}.\textit{e}., an adversary can deduce how many tuples have an identical value~\cite{DBLP:conf/ccs/NaveedKW15}; (\textit{iii}) workload-skew attack, \textit{i}.\textit{e}., an adversary, having the knowledge of frequent selection queries by observing many queries, can estimate which encrypted tuples potentially satisfy the frequent section selection queries; (\textit{iv}) access-pattern attack, \textit{i}.\textit{e}., addresses of encrypted tuples that satisfy the query. Note that computationally expensive and access-pattern-hiding cryptographic techniques (\textit{e}.\textit{g}., PIR, ORAM, DSSE, and secret-sharing) can prevent the size, frequency-count, and workload-skew attacks \emph{only} on \emph{non-skewed and non-deterministically encrypted} datasets. To the best of our knowledge, there is no cryptographic technique that prevents all the four attacks on a \emph{skewed dataset}. 
Table~\ref{tab:notations} shows notations used in this paper. 

\section{Partitioned Computation}
\label{sec:Preliminaries_and_Problem_Statements}
In this section, we first define more precisely what we mean by partitioned computing, illustrate how such a computation can leak information due to the joint processing of sensitive and non-sensitive data, discuss the corresponding security definition, and finally discuss system and adversarial models under which we will develop our solutions. 

\parskip 0pt
\setlength{\parindent}{15pt}

\smallskip
\noindent\textbf{The Partition Computation Model.}
We assume the following two entities in our model:

\noindent{\emph{A trusted database (DB) owner}} who divides a relation $R$ having attributes, say $A_1, A_2, \ldots, A_n$, into the following two relations based on row-level data sensitivity: $R_s$ and $R_{\mathit{ns}}$ containing all sensitive and non-sensitive tuples, respectively. The DB owner outsources the relation $R_{\mathit{ns}}$ to a public cloud. The tuples of the relation $R_s$ are encrypted using any existing non-deterministic encryption mechanism before outsourcing to the same public cloud. 
In our setting, the DB owner has to store metadata such as searchable values and their frequency counts, which will be used for appropriate query formulation. The DB owner is assumed to have sufficient storage for such metadata, and also computational capabilities to perform encryption and decryption. The size of metadata is smaller than the size of the original data.


  \noindent{\emph{The untrusted public cloud}} that stores the databases, executes queries, and provides answers.

\bgroup
\def\arraystretch{1}
\begin{table}[t]
\centering
\centering
\scriptsize
\begin{tabular}{|p{2cm}|p{6.8cm}|}
\hline Notations & Meaning \\ \hline
$|S|$ (or $|\mathit{NS}|$) & Number of sensitive (or non-sensitive) data values \\\hline
$R_s$ (or $R_{\mathit{ns}}$) & Sensitive (or non-sensitive) parts of a relation $R$\\\hline
$s_i$ (or $\mathit{ns}_j$)  & $i^{\mathit{th}}$ sensitive (or $j^{\mathit{th}}$ non-sensitive) value \\\hline
$\mathit{SB}$ (or $\mathit{NSB}$) & The number of sensitive (or non-sensitive) bins \\\hline
$\mathit{SB}_i$ (or $\mathit{NSB}_i$) & $i^{\mathit{th}}$ sensitive (or non-sensitive) bin \\\hline
$|\mathit{SB}|=y$ (or $|\mathit{NSB}|=x$) & Sensitive (or non-sensitive) values in a sensitive (or non-sensitive) bin or the size of a sensitive (or non-sensitive) bin\\\hline
$q(w)$ & A query, $q$, for a predicate $w$\\\hline
$q(W_{\mathit{ns}})(R_{\mathit{ns}})$ & A query, $q$, for a set, $W_{\mathit{ns}}$, of predicates in clear-text over $R_{\mathit{ns}}$\\\hline
$q(W_s)(R_s)$ & A query, $q$, for a set, $W_s$, of predicates in encrypted form over $R_s$\\\hline
$q(W)(R_s,R_{\mathit{ns}})[A]$ & A query, $q$, for a set, $W$, of values, searching on the attribute, $A$, of the relations $R_s$ and $R_{\mathit{ns}}$, where $W=W_s \cup W_{\mathit{ns}} $ \\\hline
$E(t_i)$ & $i^{\mathit{th}}$ encrypted tuple\\\hline
\end{tabular}
\caption{Notations used in the paper.}
\label{tab:notations}
\end{table}
\egroup

Let us consider a query $q$ over the relation $R$, denoted by $q(R)$. A partitioned computation strategy splits the execution of $q$ into two independent subqueries: $q(R_s)$: a query to be executed on the encrypted sensitive relation $R_s$, and $q(R_{\mathit{ns}})$: a query to be executed on the non-sensitive relation $R_{\mathit{ns}}$. The final result is computed (using a query $q_{\mathit{merge}}$) by appropriately merging the results of the two subqueries at the DB owner side. In particular, the query $q$ on a relation $R$ is partitioned, as follows: $q(R) = q_{\mathit{merge}}\Big(q(R_s), q(R_{\mathit{ns}}) \Big)$.

\begin{figure}[t]
\B
\scriptsize
\centering
\begin{tabular}{|p{.1cm}||l|l|l|l|l|l|l|}
    \hline
      &EId & FirstName & LastName & SSN  & Office & Dept \\ \hline
    $t_1$& E101 & Adam & Smith       & 111&1 & Defense  \\ \hline
    $t_2$& E259 & John & Williams   & 222&2 & Design   \\ \hline
    $t_3$& E199 & Eve  & Smith      & 333&2  & Design   \\ \hline
    $t_4$& E259 & John & Williams   & 222&6 & Defense   \\ \hline
    $t_5$& E152 & Clark & Cook   & 444&1 & Defense   \\ \hline
    $t_6$& E254 & David & Watts   & 555&4 & Design   \\ \hline
    $t_7$& E159 & Lisa & Ross   & 666&2 & Defense   \\ \hline
    $t_8$& E152 & Clark & Cook   & 444&3 & Design   \\ \hline

  \end{tabular}
  \caption{A relation: \textit{Employee}.}
  \label{fig:employee relation}
  \BBB\B
\end{figure}

\begin{figure}[h]
\begin{center}
  \begin{minipage}[t]{.3\linewidth}
  \centering
  \scriptsize
\begin{tabular}{|l|l|}\hline
         EId  & SSN   \\\hline
     E101 & 111   \\\hline
     E259 & 222   \\\hline
     E199 & 333   \\\hline
     E152 & 444   \\\hline
     E254 & 555   \\\hline
     E159 & 666   \\\hline
  \end{tabular}
  \subcaption{A sensitive relation: \textit{Employee1}.}
    \label{fig:employee1 relation}
  \end{minipage}
  \begin{minipage}[t]{.68\linewidth}
  \centering
 \scriptsize
\centering
\begin{tabular}{|p{.1cm}||p{.42cm}|p{.85cm}|p{.83cm}|l|l|l|}\hline
         & EId  & FirstName & LastName &  Office & Dept \\ \hline
    $t_1$& E101 & Adam & Smith       & 1 & Defense  \\ \hline
    $t_4$& E259 & John & Williams   & 6 & Defense   \\ \hline
    $t_5$& E152 & Clark & Cook   &1 & Defense   \\ \hline
    $t_7$& E159 & Lisa & Ross   &2 & Defense   \\ \hline
  \end{tabular}
  \subcaption{A sensitive relation: \textit{Employee2}.}
\label{fig:employee2 relation}
  \end{minipage}
  \begin{minipage}[t]{.98\linewidth}
  \centering
  \scriptsize
\centering
\begin{tabular}{|p{.1cm}||l|l|l|l|l|l|}
    \hline
         & EId  & FirstName & LastName &  Office & Dept \\ \hline
    $t_2$& E259 & John & Williams   &2 & Design   \\ \hline
    $t_3$& E199 & Eve  & Smith      &2  & Design   \\ \hline
    $t_6$& E254 & David & Watts   & 4 & Design   \\ \hline
    $t_8$& E152 & Clark & Cook   & 3 & Design   \\ \hline
  \end{tabular}
  \subcaption{A non-sensitive relation: \textit{Employee3}.}
\label{fig:employee3 relation}
  \end{minipage}
\end{center}
\BB
\caption{Three relations obtained from \textit{Employee} relation.}
\label{fig:three realtions}
\end{figure}

\noindent Let us illustrate partitioned computations through an example.

\noindent\textbf{Example 1.} Consider an \emph{Employee} relation, see Figure~\ref{fig:employee relation}. Note that the notation $t_i$ ($1\leq i\leq 8$) is not an attribute of the relation; we used this to indicate the $i^{\mathit{th}}$ tuple. In this relation, the attribute {\em SSN} is sensitive, and furthermore, all tuples of employees for the {\em Dept} $=$ ``\texttt{Defense}'' are sensitive. In such a case, the \emph{Employee} relation may be stored as the following three relations: (\textit{i}) \emph{Employee1} with attributes {\em EId} and {\em SSN} (see Figure~\ref{fig:employee1 relation}); (\textit{ii}) \emph{Employee2} with attributes {\em EId}, {\em FirstName}, {\em LastName}, {\em Office}, and {\em Dept}, where {\em Dept} $=$ ``\texttt{Defense}'' (see Figure~\ref{fig:employee2 relation}); and (\textit{iii}) \emph{Employee3} with attributes {\em EId}, {\em FirstName}, {\em LastName}, {\em Office}, and {\em Dept}, where {\em Dept} $<>$ ``\texttt{Defense}'' (see Figure~\ref{fig:employee3 relation}). Since the relations \emph{Employee1} and \emph{Employee2} (Figures~\ref{fig:employee1 relation} and~\ref{fig:employee2 relation}) contain only sensitive data, these two relations are encrypted before outsourcing, while \emph{Employee3} (Figure~\ref{fig:employee3 relation}), which contains only non-sensitive data, is outsourced in clear-text. We assume that the sensitive data is strongly encrypted such that the property of \emph{ciphertext indistinguishability} is achieved. Thus, the two occurrences of \texttt{E152} have two different ciphertexts.

%
%
%
%

Consider a query \texttt{q}: \textit{SELECT FirstName, LastName, Office, Dept from Employee where FirstName = John}. In the partitioned computation, the query \texttt{q} is partitioned into two subqueries: $q_s$ that executes on \textit{Employee2}, and $q_{\mathit{ns}}$ that executes on \textit{Employee3}. $q_s$ will retrieve the tuple $t_4$ while $q_{\mathit{ns}}$ will retrieve the tuple $t_2$. $q_{merge}$ in this example is simply a union operator. Note that the execution of the query \texttt{q} will also retrieve the same tuples. However, such a partitioned computation, if performed naively, leads to inferences about sensitive data from non-sensitive data. Before discussing inference attacks, we first present the adversarial model.

\smallskip
\noindent\textbf{Adversarial Model.}
We assume an honest-but-curious (HBC) adversary~\cite{DBLP:conf/stoc/CanettiFGN96}, which is considered in the standard setting for security in the public cloud that is \textit{not trustworthy}. An HBC adversarial public cloud stores an outsourced dataset without tampering, correctly computes assigned tasks, and returns answers; however, it may exploit side knowledge (\textit{e}.\textit{g}., query execution, background knowledge, and the output size) to gain as much information as possible about the sensitive data. 
Furthermore, the HBC adversary can eavesdrop on the communication channels between the cloud and the DB owner and that may help in gaining knowledge about sensitive data, queries, or results; hence, a secure channel is assumed. In our setting, the adversary has full access to the following:


  \noindent\emph{{All the non-sensitive data.}} For example, for the Employee relation in Example 1, an adversary knows the complete \emph{Employee3} relation (refer to Figure~\ref{fig:employee3 relation}).

  \noindent\emph{{Auxiliary/background information of the sensitive data.}} The auxiliary information~\cite{DBLP:conf/ccs/NaveedKW15,DBLP:conf/ccs/KellarisKNO16} may contain metadata, schema of the relation, and the number of tuples in the relation (note that having an adversary with the auxiliary information is also considered in literature). In Example 1, the adversary knows that there are two sensitive relations, one of them containing six tuples and the other one containing four tuples, in the \emph{Employee1} and the \emph{Employee2} relations; Figures~\ref{fig:employee1 relation} and~\ref{fig:employee2 relation}. In contrast, the adversary is not aware of the following information before the query execution: how many people work in a specific sensitive department, is a specific person working only in a sensitive department, only in a non-sensitive department, or both.

  \noindent\emph{{Adversarial view.}} When executing a query, an adversary knows which encrypted sensitive tuples and cleartext non-sensitive tuples are sent in response to a query. We refer this as the adversarial view, denoted by $\mathit{AV}$:
$\mathit{AV} = \mathit{In}_c \cup \mathit{Op}_c$, where $\mathit{In}_c$ refers to the query arrives at the cloud and $\mathit{Op}_c$ refers to the encrypted and non-encrypted tuples, transmitted in response to $\mathit{In}_c$. For example, the first row of Table~\ref{tab:answer table} shows an adversarial view that shows that $\mathit{Op}_c= t_2$ tuples from the non-sensitive relation and encrypted $\mathit{Op}_c=t_4$ tuples from the sensitive relation are returned to answer the query for $\mathit{In}_c=$ \texttt{E259}.


  \noindent\emph{{Some frequent query values.}} The adversary observes query predicates on the non-sensitive data, and hence, can deduce the most frequent query predicates by observing many queries.
\parskip 0pt
\setlength{\parindent}{15pt}

\smallskip
\noindent\textbf{Inference Attacks in Partitioned Computations.}
To see the inference attack on the sensitive data while jointly processing sensitive and non-sensitive data, consider following three queries on the \emph{Employee2} and \emph{Employee3} relations; refer to Figures~\ref{fig:employee2 relation} and~\ref{fig:employee3 relation}.

\noindent\textbf{Example 2.} (\textit{Q1}) retrieve tuples of employee \texttt{E259}, (\textit{Q2}) retrieve tuples of employee \texttt{E101}, and (\textit{Q3}) retrieve tuples of employee \texttt{E199}.\footnote{We used random \emph{Eids}, which is common in a real employee relation. In contrast, in sequential ids, the absence of an id from the non-sensitive relation directly informs the adversary that the given id exists in the sensitive relation.} When answering a query, the adversary knows the tuple ids of retrieved encrypted tuples and the full information of the returned non-sensitive tuples. We refer to this information gain by the adversary as the \emph{adversarial view}, see Table~\ref{tab:answer table}, where $\mathit{E(t_i)}$ denotes an encrypted tuple $t_i$.
\begin{table}[h]
\B
 \centering
    \begin{tabular}{|l|l|l|}
    \hline
    Query value & \multicolumn{2}{|c|}{Returned tuples/Adversarial view}           \\ \hline
    ~                          & Employee2 & Employee3 \\ \hline
    E259                      & $\mathit{E(t_4)}$        & $t_2$        \\ \hline
    E101                              & $\mathit{E(t_1)}$        & null      \\ \hline
    E199                            & null        & $t_3$      \\ \hline
    \end{tabular}
    \caption{Queries and returned tuples/adversarial view.}
    \label{tab:answer table}
    \BB
\end{table}

Outputs of the above three queries will reveal enough information to learn something about sensitive data. In \emph{Q1}, the adversary learns that \texttt{E259} works in both sensitive and non-sensitive departments, because the answers obtained from the two relations contribute to the final answer. Moreover, the adversary may learn which sensitive tuple has an \emph{Eid} equals to \texttt{E259}. In \emph{Q2}, the adversary learns that \texttt{E101} works only in a sensitive department, because the query will not return any answer from the Employee3 relation. In \emph{Q3}, the adversary learns that \texttt{E199} works only in a non-sensitive department.

%
\smallskip
\noindent\textbf{The Query Binning (QB) Approach.}
To prevent the inference attack in a partitioned computation, a new security definition is needed. Before discussing the formal definition of partitioned data security (\S\ref{subsec:Security Definition and Correctness}), we provide a possible solution to prevent inference attacks and then intuition for the security definition.

The query binning (QB) strategy stores a non-sensitive relation, say $R_{\mathit{ns}}$, in clear-text while it stores a sensitive relation, say $R_s$, using a cryptographically secure approach. QB prevents leakage such as in Example 2 by appropriately mapping a query for a predicate, say $q(w)$, to corresponding queries both over the non-sensitive relation, say $q(W_{\mathit{ns}})(R_{\mathit{ns}})$, and encrypted relation, say $q(W_s)(R_s)$, which represent a set of predicates (or selection queries) that are executed over the relation $R_{\mathit{ns}}$ in plaintext and over the sensitive relation $R_s$, using the underlying cryptographic method, respectively. The set of predicates in $q(W_{\mathit{ns}})(R_\mathit{ns})$ (likewise in $q(W_S)(R_s)$) correspond to the non-sensitive (sensitive) {\em bins} including the predicate $w$, denoted by $\mathit{NSB}$ ($\mathit{SB}$). The predicates in $q(W_s)(R_s)$ are encrypted before transmitting to the cloud.

The bins are selected such that: (\textit{i}) $w \in q(W_{\mathit{ns}})(R_{\mathit{ns}}) \cap q(W_s)(R_s)$ to ensure that all the tuples containing the predicate $w$ are retrieved, and, (\textit{ii}) joint execution of the queries $q(W_{\mathit{ns}})(R_\mathit{ns})$ and $q(W_s)(R_s)$ (hereafter, denoted by $q(W)(R_s,R_\mathit{ns})$, where $W=W_s \cup W_{\mathit{ns}}$) does not leak the predicate $w$. Results from the execution of the queries $q(W_{\mathit{ns}})(R_{\mathit{ns}})$ and $q(W_s)(R_s)$ are decrypted, possibly filtered, and merged to generate the final answer. Note that \emph{bins are created only once for all the values of a searching attribute before any query is executed}. The details of the bin formation will be discussed in \S\ref{sec:Query Bucketization}.
%
%

For answering the above-mentioned three queries, QB creates two bins on sensitive parts: $\{$\texttt{E101}, \texttt{E259}$\}$, $\{$\texttt{E152}, \texttt{E159}$\}$, and two sets on non-sensitive parts: $\{$\texttt{E259}, \texttt{E254}$\}$, $\{$\texttt{E199}, \texttt{E152}$\}$. Table~\ref{tab:answer table qb} illustrates the generated adversarial view when QB is used to answer queries as shown in Example 2. In this example, row 1 of Table~\ref{tab:answer table qb} shows that this instance of QB maps the query for \texttt{E259} to $\langle$\texttt{E259}, \texttt{E254}$\rangle$ over cleartext and to encrypted version of values for $\langle$\texttt{E259}, \texttt{E101}$\rangle$ over sensitive data. Note that simply from the generated adversarial views, the adversary cannot determine the query value $w$ (\texttt{E259} in the example) or find a value that is shared between the two sets. Thus, while answering a query, the adversary cannot learn which employee works only in defense, design, or in both.
The reason is that the desired query value, $w$, is encrypted with other encrypted values of $W_s$, and, furthermore, the query value, $w$, cannot be distinguished from many requested non-sensitive values of $W_{\mathit{ns}}$, which are in clear-text. Consequently, \emph{the adversary is unable to find an intersection of the two sets, which is the exact value}.

\begin{table}[t]
  \centering
    \begin{tabular}{|l|l|l|}
    \hline
    Query value & \multicolumn{2}{|c|}{Returned tuples/Adversarial view}           \\ \hline
    ~                         & Employee2 & Employee3 \\ \hline
    E259                      & $\mathit{E(t_4)}$, $\mathit{E(t_1)}$       & $t_2$, $t_6$  \\ \hline
    E101                      & $\mathit{E(t_4)}$, $\mathit{E(t_1)}$       & $t_3$, $t_8$  \\ \hline
    E199                      & $\mathit{E(t_4)}$, $\mathit{E(t_1)}$       & $t_3$, $t_8$  \\ \hline
    \end{tabular}
    \caption{The adversarial view when following QB.}
    \label{tab:answer table qb}
    \BBB
\end{table}



\section{Partitioned Data Security}
\label{subsec:Security Definition and Correctness}
This section formalizes the notion of {\em partitioned data security} that establishes when a partitioned computation over sensitive and non-sensitive data does not leak any sensitive information. 
We begin by first formalizing the concepts of: \emph{associated values}, \emph{associated tuples}, and \emph{relationship between counts of sensitive values}.

\parskip 0pt
\setlength{\parindent}{15pt}

\smallskip
\noindent\emph{Notations used in the definitions}. Let $t_1, t_2, \ldots, t_m$ be tuples of a sensitive relation, say $R_s$. Thus, the relation $R_s$ stores the encrypted tuples $E(t_1), E(t_2), \ldots, E(t_m)$. Let $s_1,s_2,\ldots, s_{m^{\prime}}$ be values of an attribute, say $A$, that appears in one of the sensitive tuples of $R_s$. Note that $m^{\prime} \leq m$, since several tuples may have an identical value. Furthermore, $s_i \in \mathit{Domain}(A)$, $i=1,2, \dots, m^{\prime}$, where $\mathit{Domain}(A)$ represents the domain of values the attribute $A$ can take. By $\#_s(s_i)$, we refer to the number of sensitive tuples that have $s_i$ as the value for attribute $A$. We further define $\#_s(v) = 0, \forall v \in \mathit{Domain}(A)$, $v \notin s_1,s_2,\ldots, s_{m^{\prime}}$. Let $t_1, t_2, \ldots, t_n$ be tuples of a non-sensitive relation, say $R_{\mathit{ns}}$. Let $\mathit{ns}_1,\mathit{ns}_2,\ldots, \mathit{ns}_{n^{\prime}}$ be values of the attribute $A$ that appears in one of the non-sensitive tuples of $R_{\mathit{ns}}$. In analogy with the case where the relation is sensitive, $n^{\prime} \leq n$, and $\mathit{ns}_i \in \mathit{Domain}(A)$, $i=1,2, \dots, n^{\prime}$.

\smallskip
\noindent\emph{\textbf{Associated values.}} Let $e_i = E(t_i)[A]$ be the encrypted representation of an attribute value of $A$ in a sensitive tuple of the relation $R_s$, and $\mathit{ns}_j$ be a value of the attribute $A$ for some tuple of the relation $R_{\mathit{ns}}$. We say that $e_i$ is \emph{associated} with $\mathit{ns}_j$, (denoted by $\overset{\mathrm{a}}{=}$), if the plaintext value of $e_i$ is identical to the value $\mathit{ns}_j$. In Example 1, the value of the attribute \texttt{Eid} in tuple $t_4$ (of \emph{Employee2}, see Figure~\ref{fig:employee2 relation}) is associated with the value of the attribute \texttt{Eid} in tuple $t_2$ (of \emph{Employee3}, see Figure~\ref{fig:employee3 relation}), since both values correspond to \texttt{E259}.

\smallskip
\noindent\emph{\textbf{Associated tuples.}} Let $t_i$ be a sensitive tuple of the relation $R_s$ (\textit{i}.\textit{e}., $R_s$ stores encrypted representation of $t_i$) and $t_j$ be a non-sensitive tuple of the relation $R_{\mathit{ns}}$. We state that $t_i$ is associated with $t_j$ (for an attribute, say $A$) iff the value of the attribute $A$ in $t_i$ is associated with the value of the attribute $A$ in $t_j$ (\textit{i}.\textit{e}., $t_i[A] \overset{\mathrm{a}}{=} t_j[A]$). Note that this is the same as stating that the two values of attribute $A$ are equal for both tuples.

\smallskip
\noindent\emph{\textbf{Relationship between counts of sensitive values.}} Let $v_i$ and $v_j$ be two distinct values in $\mathit{Domain}(A)$. We denote the relationship between the counts of sensitive tuples with these $A$ values (\textit{i}.\textit{e}., $\#_s(v_i)$ (or $\#_s(v_j)$)) by $v_i \overset{\mathrm{r}}{\sim} v_j$. Note that $\overset{\mathrm{r}}{\sim}$ can be one of $<, =$, or $>$ relationships. For instance, in Example 1, the \texttt{E101} $\overset{\mathrm{r}}{\sim}$ \texttt{E259} corresponds to $=$, since both values have exactly one sensitive tuple (see Figure~\ref{fig:employee2 relation}), while \texttt{E101} $\overset{\mathrm{r}}{\sim}$ \texttt{E199} is $>$, since there is one sensitive tuple with value \texttt{E101} while there is no sensitive tuple with \texttt{E199}.

Given the above definitions, we can now formally state the security requirement that ensures that simultaneous execution of queries over sensitive (encrypted) and non-sensitive (plaintext) data does not leak any information. Before that, we wish to mention the need of a new security definition in our context. The inference attack in the partitioned computing can be considered to be related to the known-plaintext attack (KPA) wherein the adversary knows some plaintext data which is hidden in a set of ciphertext. In KPA, the adversary's goal is to determine which ciphertext data is related to a given plaintext, \textit{i}.\textit{e}., determining a mapping between ciphertext and the corresponding plaintext data representing the same value. In our setup, non-sensitive values are visible to the adversary in plaintext. However, the attacks are different since, unlike the case of KPA, in our setup, the ciphertext data might not contain any data value that is the same as some non-sensitive data visible to the adversary in plaintext.\footnote{The HBC adversary cannot launch the chosen-plaintext attack (CPA) and the chosen-ciphertext attack (CCA). Since the sensitive data is non-deterministically encrypted (by our assumption), it is not prone to the ciphertext only attack (COA).}


\medskip
\noindent\textbf{Definition: Partitioned Data Security.} Let $R$ be a relation containing sensitive and non-sensitive tuples. Let $R_s$ and $R_{\mathit{ns}}$ be the sensitive and non-sensitive relations, respectively. Let $\mathit{AV}$ be an adversarial view generated for a query $q(w)(R_s,R_{\mathit{ns}})[A]$, where the query, $q$, for a value $w$ in the attribute $A$ of the $R_s$ and $R_{\mathit{ns}}$ relations. Let $X$ be the auxiliary information about the sensitive data, and $\mathit{Pr_{Adv}}$ be the probability of the adversary knowing any information. A query execution mechanism ensures the partitioned data security if the following two properties hold:

\noindent\textbf{(1)}
$\mathit{Pr}_{\mathit{Adv}}[e_i \overset{\mathrm{a}}{=} \mathit{ns}_j|X] = \mathit{Pr}_{\mathit{Adv}}[e_i \overset{\mathrm{a}}{=} \mathit{ns}_j|X, \mathit{AV}]$, where $e_i= E(t_i)[A]$ is the encrypted representation for the attribute value $A$ for any tuple $t_i$ of the relation $R_s$ and $\mathit{ns}_j$ is a value for the attribute $A$ for any tuple of the relation $R_{\mathit{ns}}$.

\noindent\textbf{(2)}
$\mathit{Pr}_{\mathit{Adv}}[v_i \overset{\mathrm{r}}{\sim} v_j|X] = \mathit{Pr}_{\mathit{Adv}}[v_i \overset{\mathrm{r}}{\sim} v_j | X, \mathit{AV}]$, for all $v_i, v_i \in \mathit{Domain}(A)$.

\parskip 0pt
\setlength{\parindent}{15pt}


\smallskip
Equation (1) captures the fact that an initial probability of \emph{associating} a sensitive tuple with a non-sensitive tuple will be identical after executing a query on the relations, \textit{i}.\textit{e}., an adversary cannot learn anything from an adversarial view generated after the query execution. Satisfying this condition also prevents an adversary to have success against KPA. Equation (2) states that the probability of an adversary gaining information about the relative frequency of sensitive values does not increase after the query execution. In Example 2, an execution of any three queries (for values \texttt{E101}, \texttt{E199}, or \texttt{E259}) without using QB does not satisfy Equation (1). For example, the query for \texttt{E199} retrieves the only tuple from non-sensitive relation, and that changes the probability of estimating whether \texttt{E199} is sensitive or non-sensitive to 0 than an initial probability of the same estimation, which was 1/4. Hence, an execution of the three queries violates partitioned data security. However, the query execution for \texttt{E259} and \texttt{E101} satisfies Equation (2), since the count of returned tuples from \emph{Employee2} is equal. Hence, the adversary cannot distinguish between the count of the values (\texttt{E259} and \texttt{E101}) in the domain of \texttt{Eid} of \emph{Employee2} relation.

%
%

\section{Query Binning Technique}
\label{sec:Query Bucketization}
We develop our strategy initially under the assumption that queries are only on a single attribute, say $A$. QB approach takes as inputs: (\textit{i}) the set of data values (of the attribute $A$) that are sensitive, along with their counts, and (\textit{ii}) the set of data values (of the attribute $A$) that are non-sensitive, along with their counts. QB returns partitions of attribute values that form the query bins for both the sensitive and for the non-sensitive parts of the query. We begin in \S\ref{subsec:The Base Case} by developing the approach for the case when a sensitive tuple is associated with at most one non-sensitive tuple (Algorithm~\ref{alg:bin_creation}). Finally, we provide a general strategy to create bins when a sensitive tuple is associated with several non-sensitive tuples, in \S\ref{subsec:Multiple Records with Multiple Values}.

\parskip 0pt
\setlength{\parindent}{15pt}

Informally, QB distributes attribute values in a matrix, where rows are sensitive bins, and columns are non-sensitive bins. For example, suppose there are 16 values, say $0, 1,\ldots,15$, and assume all the values have sensitive and associated non-sensitive tuples. Now, the DB owner arranges 16 values in a $4\times4$ matrix, as follows:
\begin{center}
\begin{tabular}{|l|l|l|l|l|}
  \hline
             & $\mathit{NSB}_0$ & $\mathit{NSB}_1$ & $\mathit{NSB}_2$ & $\mathit{NSB}_3$  \\  \hline  \hline
    $\mathit{SB}_0$ & 11  & 2 & 5 & 14 \\  \hline
    $\mathit{SB}_1$ &  10&  3&  8&7  \\  \hline
    $\mathit{SB}_2$ &  0&  15& 6 & 4 \\  \hline
    $\mathit{SB}_3$ &  13&  1& 12 &9  \\  \hline
    \end{tabular}
\end{center}

Here, we have four sensitive bins: $\mathit{SB}_0$ \{11,2,5,14\}, $\mathit{SB}_1$ \{10,3,8,7\}, $\mathit{SB}_2$ \{0,15,6,4\}, $\mathit{SB}_3$ \{13,1,12,9\}, and four non-sensitive bins: $\mathit{NSB}_0$ \{11,10,0,13\}, $\mathit{NSB}_1$ \{2,3,15,1\}, $\mathit{NSB}_2$ \{5,8,6,12\}, $\mathit{NSB}_3$ \{14,7,4,9\}. When a query arrives for a value 1, the DB owner searches for tuples containing values 2,3,15,1 (viz. $\mathit{NSB}_1$) on the non-sensitive data and values in $\mathit{SB}_3$ (viz., 13,1,12,9) on the sensitive data using a cryptographic mechanism integrated into QB. We will show that in the proposed approach, while the adversary learns that a query corresponds to one of the four values in $\mathit{NSB}_1$, since query values in $\mathit{SB}_3$ are encrypted, the adversary does not learn the actual sensitive value or the actual non-sensitive value that is identical to a clear-text sensitive value.

\subsection{The Base Case}
\label{subsec:The Base Case}
QB consists of two steps. First, query bins are created (information about which will reside at the DB owner) using which queries will be rewritten. The second step consists of rewriting the query based on the binning.
Here, QB is explained for the base case, where a sensitive tuple, $t_s$, is associated with at most a single non-sensitive tuple, $t_{\mathit{ns}}$, and vice versa (\textit{i}.\textit{e}., $\overset{\mathrm{a}}{=}$ is a 1:1 relationship). Thus, if the value has two tuples, then one of them must be sensitive and the other one must be non-sensitive, but both the tuples cannot be sensitive or non-sensitive. A value can also have only one tuple, either sensitive or non-sensitive. Note that $t_1, t_2, \ldots, t_l$ are sensitive tuples, with values of an attribute $A$ being $s_1, s_2, \ldots s_n$, $s_i$ $\neq$ $s_j$ if $i$ $\neq$ $j$. Thus, in the remainder of the section, we will refer to association between encrypted value $E(t_i)[A]$ and a non-sensitive value $\mathit{ns}_j$ simply as an association between values $s_i$ and $\mathit{ns}_j$, where $s_i$ is the cleartext representation of $E(t_i)[A]$ and $\mathit{ns}_j$ is a value in the attribute $A$ of a non-sensitive relation; \textit{i}.\textit{e}., $s_i \overset{\mathrm{a}}{=} \mathit{ns}_j$ represents $E(t_i)[A] \overset{\mathrm{a}}{=} \mathit{ns}_j$.

\parskip 0pt
\setlength{\parindent}{15pt}

The scenario in Example 1 satisfies the base case. The \emph{EId} attribute values corresponding to sensitive tuples include $\langle$\texttt{E101}, \texttt{E259}$, \texttt{E152}, \texttt{E159}\rangle$ and corresponding to non-sensitive tuples are $\langle$\texttt{E199}, \texttt{E259}, \texttt{E254}, \texttt{E152}$\rangle$ for which $\overset{\mathrm{a}}{=}$ is 1:1. We discuss QB under the above assumption, but relax the assumption in \S\ref{subsec:Multiple Records with Multiple Values}. Before describing QB, we first define the concept of {\em approximately square factors of a number}.

\smallskip
\noindent\textbf{Approximately square factors.} \emph{We say two numbers, say $x$ and $y$, are \emph{approximately square factors of a number}, say $n>0$, if $x \times y =n$, and $x$ and $y$ are equal or close to each other such that the difference between $x$ and $y$ is less than the difference between any two factors, say $x^{\prime}$ and $y^{\prime}$, of $n$ such that  $x^{\prime} \times y^{\prime} = n$.}

\smallskip
\noindent\textbf{Step 1: Bin-creation}. QB, described in Algorithm~\ref{alg:bin_creation}, finds two approximately square factors of $|\mathit{NS}|$, say $x$ and $y$, where $x\geq y$. QB creates $\mathit{SB}=x$ sensitive bins, where each sensitive bin contains at most $y$ values. Thus, we assume $|S|\geq x$. QB, further, creates $\mathit{NSB}=\lceil|NS|/x\rceil$ non-sensitive bins, where each non-sensitive bin contains at most $|\mathit{NSB}|=x$ values. Note that we are assuming that $|S|\leq|\mathit{NS}|$. (QB can also handle the case of $|S|>|\mathit{NS}|$ by applying Algorithm~\ref{alg:bin_creation} in a reverse way, \textit{i}.\textit{e}., factorizing $|S|$.)
\parskip 0pt
\setlength{\parindent}{15pt}

\noindent\emph{\textbf{Assignment of sensitive values.}} We number the sensitive bins from 0 to $x-1$ and the values therein from 0 to $y-1$. To assign a value to sensitive bins, QB first permutes the set of sensitive values. This permutation is kept secret from the adversary by the DB owner.\footnote{The DB owner permutes sensitive values to prevent the adversary to create bins at her end; \textit{e}.\textit{g}., if the adversary knows that employee ids are ordered, she can also create bins by knowing the number of resultant tuples to a query. For simplicity, we do not show permuted sensitive values in any figure.} To assign sensitive values to sensitive bins, QB takes the $i^{\mathit{th}}$ sensitive value and assigns it to the $(i$ $\mathit{modulo}$ $x)^{\mathit{th}}$ sensitive bin (see Lines~\ref{ln:permute} and~\ref{ln:sesnitive_allocate} of Algorithm~\ref{alg:bin_creation}).

\noindent\emph{\textbf{Assignment of non-sensitive values.}} We number the non-sensitive bins from 0 to $\lceil|\mathit{NS}|\rceil/x-1$ and values therein from 0 to $x-1$. To assign non-sensitive values, QB takes a sensitive bin, say $j$, and its $i^{\mathit{th}}$ sensitive value. Assign the non-sensitive value associated with the $i^{\mathit{th}}$ sensitive value to the $j^{\mathit{th}}$ position of the $i^{\mathit{th}}$ non-sensitive bin. Here, if each value of a sensitive bin has an associated non-sensitive value and $|S| = |\mathit{NS}|$, then QB has assigned all the non-sensitive values to their bins (Line~\ref{ln:assign_value_to_NS_bucket} of Algorithm~\ref{alg:bin_creation}). Note that it may be the case that only a few sensitive values have their associated non-sensitive values and $|S| \leq |\mathit{NS}|$. In this case, we assign the sensitive and their associated non-sensitive values to bins like we did in the previous case. However, we need to assign the non-sensitive values that are not associated with a sensitive value, by filling all the non-sensitive bins to size $x$ (Line~\ref{ln:assign_remaining_NS} of Algorithm~\ref{alg:bin_creation}).


\LinesNotNumbered \begin{algorithm}[!t]
\textbf{Inputs:} $|\mathit{NS}|$: the number of values in the non-sensitive data,

$|S|$: the number of values in the sensitive data.

\textbf{Outputs:} $\mathit{SB}$: sensitive bins; $\mathit{NSB}$: non-sensitive bins

\nl{\bf Function $\mathit{create\_bins(S,NS)}$} \nllabel{ln:function_create_bucket}
\Begin{
\nl Permute all sensitive values \nllabel{ln:permute}

\nl $x, y \leftarrow \mathit{approx\_sq\_factors(|NS|)}$: $x \geq y$ \nllabel{ln:largest_divisors}

\nl $|\mathit{NSB}| \leftarrow x$, $\mathit{NSB} \leftarrow \lceil |\mathit{NS}|/x\rceil$, $\mathit{SB} \leftarrow x$, $|\mathit{SB}| \leftarrow y$ \nllabel{ln:number_of_buckets}

\nl \lFor{$i \in (1,|S|)$}{$\mathit{SB}[i$ modulo $x][\ast]\leftarrow S[i]$\nllabel{ln:sesnitive_allocate}}

\nl \lFor{$(i,j)\in (0,\mathit{SB}-1),(0,|\mathit{SB}|-1)$}{$\mathit{NSB}[j][i]\leftarrow \mathit{allocateNS(\mathit{SB}[i][j])}$ \nllabel{ln:assign_value_to_NS_bucket}}

\nl \lFor{$i\in (0,\mathit{NSB}-1)$}{$\mathit{NSB}[i][\ast]\leftarrow$ fill the bin if empty with the size limit to $x$ \nllabel{ln:assign_remaining_NS}}

\nl \Return $\mathit{SB}$ and $\mathit{NSB}$
}


\nl{\bf Function $\mathit{allocateNS(\mathit{SB}[i][j])}$} \nllabel{ln:function_allocateNS}
\Begin{
find a non-sensitive value associated with the $j^{\mathit{th}}$ sensitive value of the $i^{\mathit{th}}$ sensitive bin
}

\caption{Bin-creation algorithm, the base case.}
\label{alg:bin_creation}
\end{algorithm}
\setlength{\textfloatsep}{0pt}

\noindent\emph{Aside}. Note that QB assigned at least as many values in a non-sensitive bin as it assigned to a sensitive bin. QB may form the non-sensitive and sensitive bins in such a way that the number of values in sensitive bins is higher than the non-sensitive bins. We chose sensitive bins to be smaller since the processing time on encrypted data is expected to be higher than clear-text data processing; hence, by searching and retrieving fewer sensitive tuples, we decrease the encrypted data-processing time.

\smallskip
\noindent\textbf{Step 2: Bin-retrieval -- answering queries}. Algorithm~\ref{alg:bin_retrieval} presents the pseudocode for the bin-retrieval algorithm. The algorithm, first, checks the existence of a query value in sensitive bins and/or non-sensitive bins (see Lines~\ref{ln:check_each_sbucket} and~\ref{ln:check_nsbuckets} of Algorithm~\ref{alg:bin_retrieval}). If the value exists in a sensitive bin and a non-sensitive bin, the DB owner retrieves the corresponding two bins (see Line~\ref{ln:retrieve_bin}). Note that here the adversarial view is not enough to leak the query value or to find a value that is shared between the two bins. The reason is that the desired query value is encrypted with a set of other encrypted values and, furthermore, the query value is obscured in many requested non-sensitive values, which are in clear-text. Consequently, the adversary is unable to find an intersection of the two bins, which is the exact value.


\DontPrintSemicolon
\LinesNotNumbered \begin{algorithm}[!t]
\textbf{Inputs:} $w$: the query value. \textbf{Outputs:} $\mathit{SB}_a$ and $\mathit{NSB}_b$: one sensitive bin and one non-sensitive bin to be retrieved for answering $w$.

\textbf{Variables:} $\mathit{found}\leftarrow$ \textbf{false}

\nl{\bf Function $\mathit{retrieve\_bins(q(w))}$} \nllabel{ln:function_retrieve_bucket}
\Begin{

\nl \For{$(i,j) \in (0,\mathit{SB}-1),(0,|\mathit{SB}|-1)$\nllabel{ln:check_each_sbucket}}{\If{$w=\mathit{SB}_i[j]$}{

\nl \Return $\mathit{SB}_i$ and $\mathit{NSB}_j$; $\mathit{found} \leftarrow$ \textbf{true}; \textbf{break} \nllabel{ln:retrieve_rule_svalue}
}}

\nl \If{$\mathit{found} \neq$ \textbf{\textnormal{\textbf{true}}}\nllabel{ln:check_nsbuckets}}{
\nl \For{$(i,j) \in (0,\mathit{NSB}-1),(0,|\mathit{NSB}|-1)$}{

\nl \If{$w=\mathit{NSB}_i[j]$}{\Return $\mathit{NSB}_i$ and $\mathit{SB}_j$; \textbf{break}\nllabel{ln:retrieve_rule_nsvalue}}}}

\nl Retrieve the desired tuples from the cloud by sending encrypted values of the bin $\mathit{SB}_i$ (or $\mathit{SB}_j$) and clear-text values of the bin $\mathit{NSB}_j$ (or $\mathit{NSB}_i$) to the cloud

\nllabel{ln:retrieve_bin}

}
\caption{Bin-retrieval algorithm.}
\label{alg:bin_retrieval}
\end{algorithm}
\setlength{\textfloatsep}{0pt}

\begin{figure}[!t]
\centering
\includegraphics[scale=0.5]{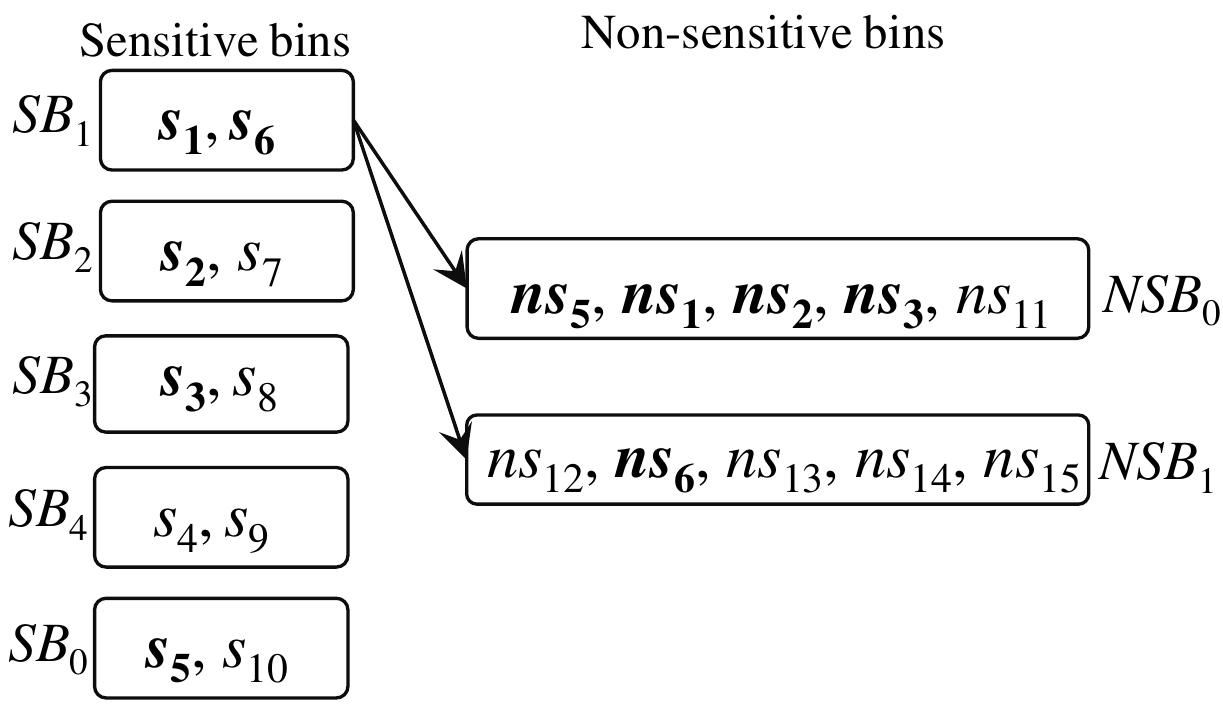}
\caption{QB for 10 sensitive and 10 non-sensitive values.}
\label{fig:qb}
\end{figure}

There are the following three other cases to consider: (\textit{i}) Some sensitive values of a bin are not associated with any non-sensitive value. For example, in Figure~\ref{fig:qb}, the sensitive values $s_4$, $s_7$, $s_8$, $s_9$, and $s_{10}$ are not associated with any non-sensitive value. (\textit{ii}) A sensitive bin does not hold any value that is associated with any non-sensitive value. For example, the sensitive bin $\mathit{SB}_4$ in Figure~\ref{fig:qb} satisfies this clause. (\textit{iii}) A non-sensitive bin containing no value that is associated with any sensitive value.

In all three cases, if the DB owner retrieves only either a sensitive or non-sensitive bin containing the value, it leads to information leakage similar to Example 2 (or incomplete answers). In order to prevent such leakage, Algorithm~\ref{alg:bin_retrieval} follows two rules stated below (see Lines~\ref{ln:retrieve_rule_svalue} and~\ref{ln:retrieve_rule_nsvalue} of Algorithm~\ref{alg:bin_retrieval}):

\noindent\textit{\textbf{Tuple retrieval rule R1.}} If the query value $w$ is a sensitive value that is at the $j^{\mathit{th}}$ position of the $i^{\mathit{th}}$ sensitive bin (\textit{i}.\textit{e}., $w=\mathit{SB}_i[j]$), then the DB owner will fetch the $i^{\mathit{th}}$ sensitive and the $j^{\mathit{th}}$ non-sensitive bins (see Line~\ref{ln:retrieve_rule_svalue} of Algorithm~\ref{alg:bin_retrieval}). By Line~\ref{ln:check_each_sbucket} of Algorithm~\ref{alg:bin_retrieval}, the DB owner knows that the value $w$ is either sensitive or non-sensitive.

\noindent\textit{\textbf{Tuple retrieval rule R2.}} If the query value $w$ is a non-sensitive value that is at the $j^{\mathit{th}}$ position of the $i^{\mathit{th}}$ non-sensitive bin, then the DB owner will fetch the $i^{\mathit{th}}$ non-sensitive and the $j^{\mathit{th}}$ sensitive bins (see Line~\ref{ln:retrieve_rule_nsvalue} of Algorithm~\ref{alg:bin_retrieval}).

Note that if query value $w$ is in both sensitive and non-sensitive bins, then both the rules are applicable, and they retrieve \emph{exactly the same} bins. In addition, if the value $w$ is neither in a sensitive or a non-sensitive bin, then there is no need to retrieve any bin.

\noindent\emph{Aside}. After knowing the bins, the DB owner sends all the sensitive values in the encrypted form and the non-sensitive values in clear-text to the cloud. The tuple retrieval based on the encrypted values reveals only the tuple addresses that satisfy the requested values. We can also hide the access-patterns by using PIR, ORAM, or DSSE on each required sensitive value. As mentioned in \S\ref{sec:introduction}, access-pattern-hiding techniques are prone to size and workload-skew attacks. Nonetheless, the use of QB with access-pattern-hiding techniques makes them secure against these attacks, which is discussed in detail in the full version. QB is designed as a general mechanism that provides partitioned data security when coupled with any cryptographic technique. For special cryptographic techniques that hide access-patterns, it may be possible to design a different mechanism that may provide partitioned data security.

\noindent
\textbf{Associated bins.} \emph{We say a sensitive bin is associated with a non-sensitive bin, if the two bins are retrieved for answering at least one query.}

Our aim when answering queries for all the sensitive and non-sensitive values using Algorithm~\ref{alg:bin_retrieval} is to associate each sensitive bin with each non-sensitive bin; resulting in the adversary being unable to predict which (if any) is the value shared between two bins.


\smallskip
\noindent\textnormal{\textbf{Example 3: QB example Step 1: Bin Creation.}} We show the bin-creation algorithm for 10 sensitive values and 10 non-sensitive values. We assume that only five sensitive values, say $s_1, s_2, s_3, s_5, s_6$, have their associated non-sensitive values, say $\mathit{ns}_1, \mathit{ns}_2,\mathit{ns}_3,\mathit{ns}_5, \mathit{ns}_{6}$, and the remaining 5 sensitive (say, $s_4, s_7,s_8, \ldots s_{10}$) and 5 non-sensitive values (say, $\mathit{ns}_{11}, \mathit{ns}_{12},\ldots, \mathit{ns}_{15}$) are not associated. For simplicity, we use different indexes for non-associated values.

QB creates 2 non-sensitive bins and 5 sensitive bins, and divides 10 sensitive values over the following 5 sensitive bins: $\mathit{SB}_0$ $\{s_5, s_{10}\}$, $\mathit{SB}_1$ $\{s_1, s_6\}$, $\mathit{SB}_2$ $\{s_2, s_7\}$, $\mathit{SB}_3$ $\{s_3, s_8\}$, $\mathit{SB}_4$ $\{s_4, s_9\}$; see Figure~\ref{fig:qb}. Now, QB distributes non-sensitive values associated with the sensitive values over two non-sensitive bins, resulting in the bin $\mathit{NSB}_0$ $\{\mathit{ns}_5,\mathit{ns}_1,\mathit{ns}_2,\mathit{ns}_3,\ast \}$ and $\mathit{NSB}_1$ $\{\ast,\mathit{ns}_6,\ast,\ast,\ast\}$, where a $\ast$ shows an empty position in the bin. In the sequel, QB needs to fill the non-sensitive bins with the remaining 5 non-sensitive values; hence, $\mathit{ns}_{11}$ is assigned to the last position of the bin $\mathit{NSB}_0$, and the bin $\mathit{NSB}_1$ contains the remaining 4 non-sensitive values such as $\{\mathit{ns}_{12},\mathit{ns}_6,\mathit{ns}_{13},\mathit{ns}_{14},\mathit{ns}_{15}\}$.

\smallskip\noindent\textbf{Example 3: QB example (continued) Step 2: Bin-retrieval.} We show how to retrieve tuples. If a query is for the sensitive value $s_2$ (see Figure~\ref{fig:qb}), then the DB owner fetches two bins $\mathit{SB}_2$ and $\mathit{NSB}_0$. If a query is for the non-sensitive value $\mathit{ns}_{13}$ or sensitive value $s_7$, then the DB owner fetches two bins $\mathit{SB}_2$ and $\mathit{NSB}_1$. Thus, it is impossible for the adversary to find (by observing the adversarial view) which is an exact query value from the non-sensitive bin and which is the sensitive value associated with one of the non-sensitive values. This fact is also clear from Table~\ref{tab:answer table_qb}, which shows that the adversarial view is not enough to leak information from the joint processing of sensitive and non-sensitive data, unlike Example 2. In Table~\ref{tab:answer table_qb}, $E(s_i)$ shows the encrypted value of $s_i$, and we are showing the adversarial view only for queries for $s_2$, $s_7$, and $\mathit{ns}_{13}$, due to space restriction. In this example, note that the bin $\mathit{SB}_2$ gets associated with both the non-sensitive bins $\mathit{NSB}_0$ and $\mathit{NSB}_1$, due to following Algorithm~\ref{alg:bin_retrieval}.
\begin{table}[h]
\scriptsize
  \centering
    \begin{tabular}{|p{1.8cm}|l|l|}
    \hline
    Exact query value & \multicolumn{2}{|c|}{Returned tuples/Adversarial view}           \\ \hline
    ~                  & Sensitive bin and data        & Non-sensitive bin and data  \\ \hline

    $s_2$ or $\mathit{ns}_2$ & $\mathit{SB}_2$\textbf{:}$\mathit{E(s_2)}$,$\mathit{E(s_7)}$ & $\mathit{NSB}_0$\textbf{:}$\mathit{ns}_1$,$\mathit{ns}_2$,$\mathit{ns}_3$,$\mathit{ns}_5$,$\mathit{ns}_{11}$ \\\hline

    $s_7$ & $\mathit{SB}_2$\textbf{:}$\mathit{E(s_2)}$,$\mathit{E(s_7)}$ & $\mathit{NSB}_1$\textbf{:}$\mathit{ns}_6$,$\mathit{ns}_{12}$,$\mathit{ns}_{13}$,$\mathit{ns}_{14}$,$\mathit{ns}_{15}$ \\\hline

    $\mathit{ns}_{13}$ & $\mathit{SB}_2$\textbf{:}$\mathit{E(s_2)}$,$\mathit{E(s_7)}$ & $\mathit{NSB}_1$\textbf{:}$\mathit{ns}_6$,$\mathit{ns}_{12}$,$\mathit{ns}_{13}$,$\mathit{ns}_{14}$,$\mathit{ns}_{15}$ \\\hline

    \end{tabular}
    \caption{Queries and returned tuples/adversarial view after retrieving tuples according to Algorithm~\ref{alg:bin_retrieval}.}
    \label{tab:answer table_qb}
\end{table}

\noindent\textbf{Algorithm Correctness.}
We will prove that QB does not lead to information leakage through the joint processing of sensitive and non-sensitive data. To prove correctness, we first define the concept of \emph{surviving matches}. Informally, we show that QB maintains surviving matches among all sensitive and non-sensitive values, resulting in all sensitive bins being associated with all non-sensitive bins. Thus, an initial condition: a sensitive value is assumed to have an identical value to one of the non-sensitive value is preserved.
\parskip 0pt
\setlength{\parindent}{15pt}

\noindent\textbf{\textit{Surviving matches.}} We define surviving matches, which are classified as either \emph{surviving matches of values} or \emph{surviving matches of bins}, as follows:

\noindent\emph{Before query execution}. Before retrieving any tuple, having an assumption that only the DB owner can decrypt an encrypted sensitive value, $E(s_i)$, the adversary cannot learn which non-sensitive value is associated with the value $s_i$. Thus, the adversary will consider that $E(s_i)$ is associated with one of the non-sensitive values. Based on this fact, the adversary can create a complete bipartite graph having $|S|$ nodes on one side and $|\mathit{NS}|$ nodes on the other side. The edges in the graph are called \emph{surviving matches of the values}. For example, before executing any query, the adversary can create a bipartite graph for 10 sensitive and 10 non-sensitive values.

\begin{figure}[t]
\begin{center}
  \begin{minipage}[t]{.45\linewidth}
  \centering
  \includegraphics[scale=0.55]{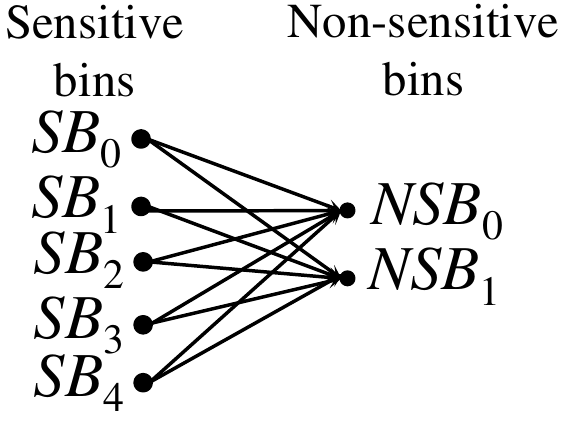}
  \subcaption{Surviving matches after the tuple retrieval following Algorithm~\ref{alg:bin_retrieval}.}
  \label{fig:survival_matching2}
  \end{minipage}
  \begin{minipage}[t]{.53\linewidth}
  \centering
  \includegraphics[scale=0.55]{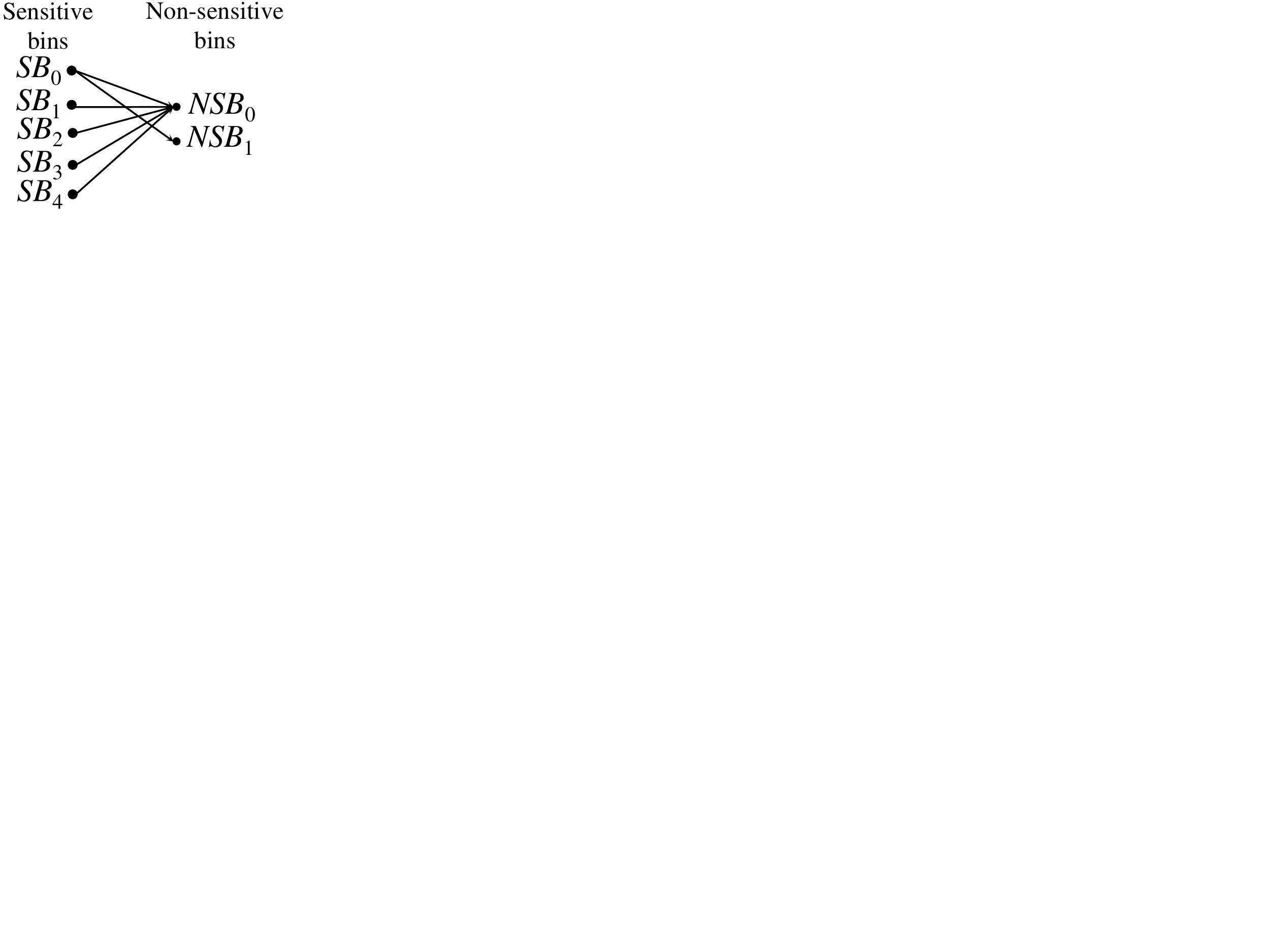}
  \subcaption{Surviving matches without following Algorithm~\ref{alg:bin_retrieval} for $\mathit{ns}_{12}$, $\mathit{ns}_{13}$, $\mathit{ns}_{14}$, $\mathit{ns}_{15}$; also see Table~\ref{tab:answer table_qb}.}
  \label{fig:survival_matching3}
  \end{minipage}
\end{center}
\caption{An example to show security of QB using surviving matches for 10 sensitive and 10 non-sensitive values.}
\label{fig:survival_mathcings}
\end{figure}

\noindent\emph{After query execution}. Recall that the query execution on the datasets creates an adversarial view that guides the adversary to create a (new) bipartite graph containing $\mathit{SB}$ nodes on one side and $\mathit{NSB}$ nodes on the other side. The edges in the new graph (obtained after the query execution) are called \emph{surviving matches of the bins}. \textit{E}.\textit{g}., after executing queries according to Algorithm~\ref{alg:bin_retrieval}, an adversary can create a bipartite graph having 5 nodes on one side and 2 nodes on another side (Figure~\ref{fig:survival_matching2}). Note that since bins contain values, the surviving matches of the bins can lead to the surviving matches of the values. Hence, from Figure~\ref{fig:survival_matching2}, the adversary can also create a bipartite graph for 10 sensitive and 10 non-sensitive values.

We show that a technique for retrieving tuples that drops some surviving matches of the bins leading to drop of the surviving matches of the values is not secure, and hence, results in the information leakage through non-sensitive data.

\smallskip
\noindent\textbf{Example 4: Dropping surviving matches.} In Figure~\ref{fig:qb}, for answering queries for associated values $s_1$, $s_2$, $s_3$, $s_5$, $s_6$, $\mathit{ns}_1$, $\mathit{ns}_2$, $\mathit{ns}_3$, $\mathit{ns}_5$, or $\mathit{ns}_6$, the DB owner must follow Line~\ref{ln:retrieve_rule_svalue} or~\ref{ln:retrieve_rule_nsvalue} of Algorithm~\ref{alg:bin_retrieval} for retrieving the two bins holding corresponding sensitive and non-sensitive data; otherwise, she cannot retrieve two bins that share a common value. 
However, for answering values $s_4$, $s_7$, $s_8$, $s_9$, $s_{10}$, $\mathit{ns}_6$, $\mathit{ns}_{12}$, $\mathit{ns}_{13}$, $\mathit{ns}_{14}$, or $\mathit{ns}_{15}$ (recall that these values are not associated), if the DB owner does not follow Algorithm~\ref{alg:bin_retrieval} and retrieves the bin containing the desired value with any randomly selected bin of the other side, then it could result in the following adversarial view; see Table~\ref{tab:wrong_view}. (We show the case when $\mathit{NSB}_1$ is only associated with bin $\mathit{SB}_1$, and bins $\mathit{SB}_2$ is only associated with bin $\mathit{NSB}_0$, since
Algorithm~\ref{alg:bin_retrieval} is not followed.)

Having such an adversarial view (Table~\ref{tab:wrong_view}), the adversary can learn the following fact that: (\textit{i}) Encrypted sensitive tuples of the bin $\mathit{SB}_2$ have associated non-sensitive tuples only in the bin $\mathit{NSB}_0$, not in $\mathit{NSB}_1$ (Figure~\ref{fig:survival_matching3}). (\textit{ii})
Non-sensitive tuples of the bin $\mathit{NSB}_1$ have their associated sensitive tuples only in the bin $\mathit{SB}_1$ (see Figure~\ref{fig:survival_matching3}). Based on this adversarial view (Table~\ref{tab:wrong_view}), the bipartite graph drops some surviving matches of the bins (see Figure~\ref{fig:survival_matching3}). However, note that in Table~\ref{tab:answer table_qb}, bin $\mathit{SB}_2$ was associated with both non-sensitive bins. Hence, a random retrieval of bins is not secure to prevent information leakage through non-sensitive data accessing.

\begin{table}[t]

\scriptsize
  \centering
    \begin{tabular}{|p{1.8cm}|l|l|}
    \hline
    Exact query value & \multicolumn{2}{|c|}{Returned tuples/Adversarial view}           \\ \hline
    ~                  & Sensitive bin and data        & Non-sensitive bin and data  \\ \hline
%
    $s_2$ or $\mathit{ns}_2$ & $\mathit{SB}_2$\textbf{:}$\mathit{E(s_2)}$,$\mathit{E(s_7)}$ & $\mathit{NSB}_0$\textbf{:}$\mathit{ns}_1$,$\mathit{ns}_2$,$\mathit{ns}_3$,$\mathit{ns}_5$,$\mathit{ns}_{11}$ \\\hline
%







    $s_6$ or $\mathit{ns}_6$ & $\mathit{SB}_1$\textbf{:}$\mathit{E(s_1)}$,$\mathit{E(s_6)}$ & $\mathit{NSB}_1$\textbf{:}$\mathit{ns}_6$,$\mathit{ns}_{12}$,$\mathit{ns}_{13}$,$\mathit{ns}_{14}$,$\mathit{ns}_{15}$ \\\hline

    $s_7$ & $\mathit{SB}_2$\textbf{:}$E(s_2)$,$E(s_7)$ &
    $\mathit{NSB}_0$\textbf{:}$\mathit{ns}_1$,$\mathit{ns}_2$,$\mathit{ns}_3$,$\mathit{ns}_5$,$\mathit{ns}_{11}$ \\\hline

    $\mathit{ns}_{12}$ & $\mathit{SB}_1$\textbf{:}$E(s_1)$,$E(s_6)$ & $\mathit{NSB}_1$\textbf{:}$\mathit{ns}_6$,$\mathit{ns}_{12}$,$\mathit{ns}_{13}$,$\mathit{ns}_{14}$,$\mathit{ns}_{15}$ \\\hline

    $\mathit{ns}_{13}$ & $\mathit{SB}_1$\textbf{:}$E(s_1)$,$E(s_6)$ & $\mathit{NSB}_1$\textbf{:}$\mathit{ns}_6$,$\mathit{ns}_{12}$,$\mathit{ns}_{13}$,$\mathit{ns}_{14}$,$\mathit{ns}_{15}$ \\\hline

    $\mathit{ns}_{14}$ & $\mathit{SB}_1$\textbf{:}$E(s_1)$,$E(s_6)$ & $\mathit{NSB}_1$\textbf{:}$\mathit{ns}_6$,$\mathit{ns}_{12}$,$\mathit{ns}_{13}$,$\mathit{ns}_{14}$,$\mathit{ns}_{15}$ \\\hline

    $\mathit{ns}_{15}$ & $\mathit{SB}_1$\textbf{:}$E(s_1)$,$E(s_6)$ & $\mathit{NSB}_1$\textbf{:}$\mathit{ns}_6$,$\mathit{ns}_{12}$,$\mathit{ns}_{13}$,$\mathit{ns}_{14}$,$\mathit{ns}_{15}$ \\\hline

    \end{tabular}
    \caption{Queries and returned tuples/adversarial view without following Algorithm~\ref{alg:bin_retrieval}.}
    \label{tab:wrong_view}
\end{table}

However, if the DB owner uses Line~\ref{ln:retrieve_rule_svalue} or~\ref{ln:retrieve_rule_nsvalue} of Algorithm~\ref{alg:bin_retrieval} for retrieving values that are not associated, the above-mentioned facts
no longer hold. Figure~\ref{fig:survival_matching2} shows each sensitive bin is associated with each non-sensitive bin, if Algorithm~\ref{alg:bin_retrieval} is followed. Thus, all the surviving matches of the bins and values are preserved after answering queries. Hence, for the example of 10 sensitive and 10 non-sensitive values, QB (Algorithms~\ref{alg:bin_creation} and~\ref{alg:bin_retrieval}) is secure, so the adversary cannot find an exact association between sensitive and  non-sensitive values.

\smallskip
\noindent\textbf{Informal security proof sketch} (see~\cite{TR2018} for a detailed proof). Let $v_1$, $v_2$, $v_3$, and $v_4$ be values containing only one sensitive and one non-sensitive tuple. Let $E_1$, $E_2$, $E_3$, and $E_4$ be encrypted representations of these values in an arbitrary order, \textit{i}.\textit{e}., it is not mandatory that $E_1$ be the encrypted representation of $v_1$. In this example, the cloud stores an encrypted relation, say $R_s$, containing four encrypted tuples with encrypted representations $E_1$, $E_2$, $E_3$, $E_4$ and a clear-text relation, say $R_{\mathit{ns}}$, containing four clear-text tuples with values $v_1$, $v_2$, $v_3$, $v_4$. The objective of the adversary is to deduce a clear-text value corresponding to an encrypted value. Note that before executing a query, the probability of an encrypted value, say $E_i$, to have the clear-text value, say $v_i$, $1\leq i\leq 4$ is 1/4, which QB maintains at the end of a query. Assume that the user wishes to retrieve the tuple containing $v_1$. By following QB, the user asks a query, say $q(E_1,E_3)(R_s)$, on the encrypted relation $R_s$ for $E_1$, $E_3$, and a query, say $q(v_1,v_2)(R_{\mathit{ns}})$, on the clear-text relation $R_{\mathit{ns}}$ for $v_1,v_2$. Here, we need that the probability of finding the clear-text value of an encrypted representation, say $E_i$, $1\leq i \leq 4$, remains identical before and after a query to satisfy the first condition of the partitioned data security. In short, the retrieval of the four tuples containing one of the following: $\langle E_1,E_3,v_1,v_2\rangle$, results in 16 possible allocations of the values $v_1$, $v_2$, $v_3$, and $v_4$ to $E_1$, $E_2$, $E_3$, and $E_4$, of which only four possible allocations have $v_1$ as the clear-text representation of $E_1$. This results in the probability of finding $E_1=v_1$ is 1/4. A similar argument also holds for other encrypted values. Hence, an initial probability of associating a sensitive value with a non-sensitive value remains identical to after executing a query.



\smallskip\noindent\textbf{A Simple Extension of the Base Case.} Algorithm~\ref{alg:bin_creation} creates bins when the number of non-sensitive data values is not a prime number, by finding the two approximately square factors. However, Algorithm~\ref{alg:bin_creation} may exhibit a relatively higher \emph{cost} (\textit{i}.\textit{e}., the number of the retrieved tuple) when the sum of the approximately square factors is high. For example, if there are 41 sensitive data values and 82 non-sensitive data values, then Algorithm~\ref{alg:bin_creation} creates 2 non-sensitive bins having 41 values in each and 41 sensitive bins having exactly one value in each. We handle the case when the number of non-sensitive values is close to a square number. We find the cost using Algorithm~\ref{alg:bin_creation}; in addition, we find a square number closest to the non-sensitive values (here 81 is the closest square number to 82) and the cost. Then, we use Algorithm~\ref{alg:bin_creation} that creates bins using a method that results in fewer retrieved tuples. In this example, we create 9 non-sensitive and 9 sensitive bins.


\subsection{General Case: Multiple Values with Multiple Tuples}
\label{subsec:Multiple Records with Multiple Values}
This section generalizes Algorithm~\ref{alg:bin_creation} to consider a case
when different data values have different numbers of associated tuples. First, we will show that sensitive values with different numbers of tuples may provide enough information to the adversary leading to the size, frequency-count attacks, and may disclose some information about the sensitive data. Hence, in the case of multiple values with multiple tuples, Algorithm~\ref{alg:bin_creation} cannot be directly implemented. We, thus, develop a strategy to overcome such a situation.

\parskip 0pt
\setlength{\parindent}{15pt}

\noindent
\textbf{Size attack scenario in the base QB.} Consider an assignment of 10 sensitive and 10 non-sensitive values to bins using Algorithm~\ref{alg:bin_creation}; see Figure~\ref{fig:qb}. Assume that a sensitive value, say $s_1$, has 1000 sensitive tuples and an associated non-sensitive value, say $\mathit{ns}_1$, has 2000 tuples, while all the other values have only one tuple each. Further, assume that \emph{each data value represents the salary of employees}. In this example, consider a query execution for a value, say $\mathit{ns}_1$. The DB owner retrieves tuples from two bins: $\mathit{SB}_1$ (containing encrypted tuples of values $s_1$ and $s_6$) and $\mathit{NSB}_0$ (containing tuples of values $\mathit{ns}_1, \mathit{ns}_2,\mathit{ns}_3,\mathit{ns}_5,\mathit{ns}_{11}$); see Figure~\ref{fig:qb}. The number of retrieved tuples satisfying the values of the bins $\mathit{SB}_1$ and $\mathit{NSB}_0$ will be highest (\textit{i}.\textit{e}., 3005) than the number of tuples retrieved based on any two other bins. Thus, the retrieval of the two bins $\mathit{SB}_1$ and $\mathit{NSB}_0$ provides enough information to the adversary to determine which one is the sensitive bin associated with the bin holding the value $\mathit{ns}_1$. Moreover, after observing many queries and having background knowledge, the adversary may estimate that 1000 people in the sensitive relation earn a salary equal to the value $\mathit{ns}_1$.

Thus, in the case of different sensitive values having different numbers of tuples, Algorithm~\ref{alg:bin_creation} cannot satisfy the \emph{second condition of partitioned data security} (\textit{i}.\textit{e}., the adversary is able to distinguish two sensitive values based on the number of retrieved tuples, which was not possible before the query execution, and concludes that a sensitive value ($s_1$ in the above example) has more tuples than any other sensitive value) though preserving all surviving matches.

\emph{In order for the second condition of partitioned data security to hold (and for the scheme to be resilient to the size and frequency-count attacks, as illustrated above), sensitive bins need to hold identical numbers of tuples}. A trivial way of doing this is to outsource some encrypted fake tuples such that the number of tuples in each sensitive bin will be identical. However, we need to be careful; otherwise, adding fake tuples in each sensitive bin may increase the \emph{cost}, if all the heavy-hitter sensitive values are allocated to a single bin. This fact will be clear in the following example.

\begin{figure}[h]
\begin{center}
  \begin{minipage}{.48\linewidth}
  \centering
  \includegraphics[scale=0.55]{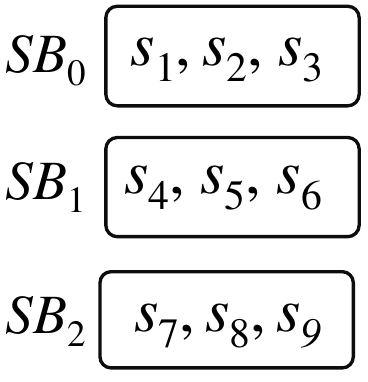}
  \B
  \subcaption{The first way.}
  \label{fig:fig_allocation_problem1}
  \end{minipage}
  \begin{minipage}{.49\linewidth}
  \centering
  \includegraphics[scale=0.55]{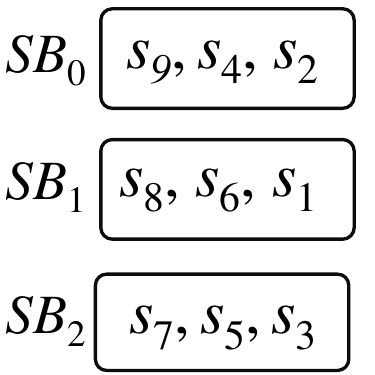}
  \B
  \subcaption{The second way.}
  \label{fig:fig_allocation_problem2}
  \end{minipage}
\end{center}
\caption{An assignment of 9 sensitive values to 3 bins.}
\label{fig:allocation_problem}
\end{figure}
\noindent\textnormal{\textbf{Example 5: (Illustrating ways to assign \emph{sensitive} values to bins to minimize the addition of fake tuples).}} Consider 9 sensitive values, say $s_1, s_2, \ldots, s_9$, having 10, 20, 30, 40, 50, 60, 70, 80, and 90 tuples, respectively (we assume that there are 9 non-sensitive values, and computed that we need 3 sensitive and 3 non-sensitive bins). There are multiple ways of assigning these values to three bins so that we need to add a minimum number of fake tuples to each bin. Figure~\ref{fig:allocation_problem} shows two different ways to assign these values to bins. Figure~\ref{fig:fig_allocation_problem2} shows the best way -- to minimize the addition of fake encrypted tuples; hence minimizing the cost. However, bins in Figure~\ref{fig:fig_allocation_problem1} require us to add 180 and 90 fake encrypted tuples to the bins $\mathit{SB}_0$ and $\mathit{SB}_1$, respectively.

There is no need to add any fake tuple if the non-sensitive values have identical numbers of tuples, because an adversary cannot deduce which sensitive bin contains sensitive tuples associated with a non-sensitive value. However, it is obvious that fake non-sensitive tuples cannot be added in clear-text.



\noindent
\textbf{Adding fake encrypted tuples.} As an assumption, we know the number of sensitive bins, say $\mathit{SB}$, using Algorithm~\ref{alg:bin_creation}. Our objective is to assign sensitive values to bins such that each bin holds identical numbers of tuples while minimizing the number of fake tuples in each bin. To do this, the strategy is as follows: (\textit{i}) Sort all the values in a decreasing order of the number of tuples. (\textit{ii}) Select $\mathit{SB}$ largest values and allocate one in each bin.
(\textit{iii}) Select the next value and find a bin that is containing the fewest number of tuples. If the bin is holding less than $y$ values, then add the value to the bin; otherwise, select another bin with the fewest number of tuples. Repeat this step, for allocating all the values to sensitive bins.
(\textit{iv}) Add fake tuples' values to the bins so that each bin contains identical numbers of tuples. (\textit{v}) Allocate non-sensitive values as per Algorithm~\ref{alg:bin_creation} (Lines~\ref{ln:assign_value_to_NS_bucket} and~\ref{ln:assign_remaining_NS}).




\section{Performance Evaluation of QB}
\label{sec:Experiments}
This section explores how effective is QB in scaling expensive cryptographic techniques by eliminating the necessity of encrypted data processing over non-sensitive data. Note that while QB prevents expensive cryptographic operations, it, nonetheless, comes with an overhead of additional search as well as communication costs.
Thus, a natural question is when the tradeoff offered by QB improves performance. We note that if a cryptographic mechanism is extremely inexpensive, \textit{e}.\textit{g}., deterministic, order-preserving encryptions, or index-based mechanisms~\cite{DBLP:conf/dbsec/ShmueliWEG05,arx-popa-2017}, 
 the overhead of QB would not justify the reduced encrypted data processing. But QB has not been designed for such techniques. Instead, our goal for QB is to couple it with techniques such as homomorphic encryptions, DPF~\cite{DBLP:conf/eurocrypt/GilboaI14}, or secret-sharing~\cite{DBLP:journals/isci/EmekciMAA14} 
that offer strong security but do not scale. Below we develop an analytical model to compare performance of cryptographic mechanisms with/without QB. We then conduct an experimental validation of the model and study QB under different choices of parameters.

\subsection{Performance Modeling of QB}
\label{subsec:Performance Modeling of QB}
For our model, we will need the following notations: (\textit{i}) $C_{\mathit{com}}$: Communication cost of moving one tuple over the network. (\textit{ii}) $C_p$ (or $C_e$): Processing cost of a single selection query on plaintext (or encrypted data). In addition, we define three parameters: (\textit{i}) $\alpha$: is the ratio between the sizes of the sensitive data (denoted by $S$) and the entire dataset (denoted by $S+\mathit{NS}$, where $\mathit{NS}$ is non-sensitive data). (\textit{ii}) $\beta$: is the ratio between the predicate search time on encrypted data using a cryptographic technique and on clear-text data. The parameter $\beta$ captures the overhead of a cryptographic technique. Note that $\beta = C_e/C_p$. 
(\textit{iii}) $\gamma$: is the ratio between the processing time of a single selection query on encrypted data and the time to transmit the single tuple over the network from the cloud to the DB owner. Note that $\gamma = C_e/C_{\mathit{com}}$. 

Based on the above parameters, we can compute the cost of cryptographic and non-cryptographic selection operations as follows: (\textit{i}) $\mathit{Cost}_{\mathit{plain}}(x, D)$: is the sum the processing cost of $x$ selection queries on plaintext data and the communication cost of moving all the tuples having $x$ predicates from the cloud to the DB owner, \textit{i}.\textit{e}., $x(\log(D)P_p+ \rho D C_{\mathit{com}})$. (\textit{ii}) $\mathit{Cost}_{\mathit{crypt}}(x, D)$: is the sum the processing cost of $x$ selection queries on encrypted data and the communication cost of moving all the tuples having $x$ predicates from the cloud to the DB owner, \textit{i}.\textit{e}., $P_e D+ \rho x D C_{\mathit{com}}$, where $\rho$ is the selectivity of the query. Note that cost of evaluating $x$ queries over encrypted data using techniques such as~\cite{DBLP:conf/sp/SongWP00,DBLP:conf/eurocrypt/GilboaI14,DBLP:journals/isci/EmekciMAA14}, is amortized and can be performed using a single scan of data. Hence, $x$ is not the factor in the cost corresponding to encrypted data processing.

Given the above, we define a parameter $\eta$ that is the ratio between the computation and communication cost of searching using QB and the computation and communication cost of searching when the entire data (viz. sensitive and non-sensitive data) is fully encrypted using the cryptographic mechanism.

{\scriptsize
$$\eta = \frac{\mathit{Cost}_{\mathit{crypt}}(|\mathit{SB}|, S)}{\mathit{Cost}_{\mathit{crypt}}(1,D)} + \frac{\mathit{Cost}_{\mathit{plain}}(|\mathit{NSB}|, \mathit{NS})}{\mathit{Cost}_{\mathit{crypt}}(1,D)}$$}
\noindent Filling out the values from above, the ratio is:
{\scriptsize $$\eta=\frac{C_e S + |\mathit{SB}|\rho D C_{\mathit{com}}}{C_e D + \rho D C_{\mathit{com}}} + \frac{|\mathit{NSB}|\log(D)C_p  + |\mathit{NSB}| \rho D C_{\mathit{com}}}{C_e D + \rho D C_{\mathit{com}}}$$}
\noindent Separating out the communication and processing costs,
{\scriptsize $$\eta = \frac{S}{D}\frac{C_e}{C_e + \rho C_{\mathit{com}}} + \frac{|\mathit{NSB}|\log(D)C_p}{C_e D + \rho D C_{\mathit{com}}}
+\frac{\rho D C_{\mathit{com}}(|\mathit{NSB}|+|\mathit{SB}|)}{C_e D + \rho D C_{\mathit{com}}}$$}
\noindent Substituting for various terms and cancelling common terms:
{\scriptsize $$\eta = \alpha \frac{1}{(1 + \frac{\rho}{\gamma})} + \frac{\log(D)}{D} \frac{|\mathit{NSB}|}{\beta (1 + \frac{\rho}{\gamma})} + \frac{\rho}{\gamma} \frac{|\mathit{NSB}| + |\mathit{SB}|}{(1 + \frac{\rho}{\gamma})}$$}
\noindent Note that $\rho/\gamma$ is very small, thus the term $(1 + \rho/\gamma)$ can be substituted by $1$. Given the above, the equation becomes: $\eta = \alpha  + \log(D)|\mathit{NSB}/D\beta + \rho(|\mathit{NSB}| + |\mathit{SB}|)/\gamma$. Note that the term $\log(D)|\mathit{NSB}|/D\beta$ is very small since $|\mathit{NSB}|$ is the number of distinct values (approx. equal to $\sqrt{|\mathit{NS}|}$) in a non-sensitive bin, while
$D$, which is the size of a database, is a large number, and $\beta$ value is also very large. Thus, the equation becomes: $\eta = \alpha  + \rho(|\mathit{SB}|+ |\mathit{NSB}|)/\gamma$.

QB is better than a cryptographic approach when $\eta < 1$, \textit{i}.\textit{e}., $\alpha  + \rho(|\mathit{SB}|+ |\mathit{NSB}|)/\gamma < 1$. Thus, $\alpha   < 1 -  \frac{\rho(|\mathit{SB}|+ |\mathit{NSB}|)}{\gamma}$. Note that the values of $|\mathit{SB}| $ and $|\mathit{NSB}|$ are $\approx\sqrt{|\mathit{NS}|}$, we can simplify the above equation to: $\alpha   < 1 -  2 \rho \sqrt{|\mathit{NS}|}/\gamma$. If we estimate $\rho$ to be roughly $1/|\mathit{NS}|$ (\textit{i}.\textit{e}., we assume uniform distribution), the above equation becomes: $\alpha  < 1 -  2/\gamma \sqrt{|\mathit{NS}|}$.

\begin{figure}[t]
\B
		\begin{center}
			\begin{minipage}[b]{.9\linewidth}
				\centering
				\includegraphics[scale=0.49]{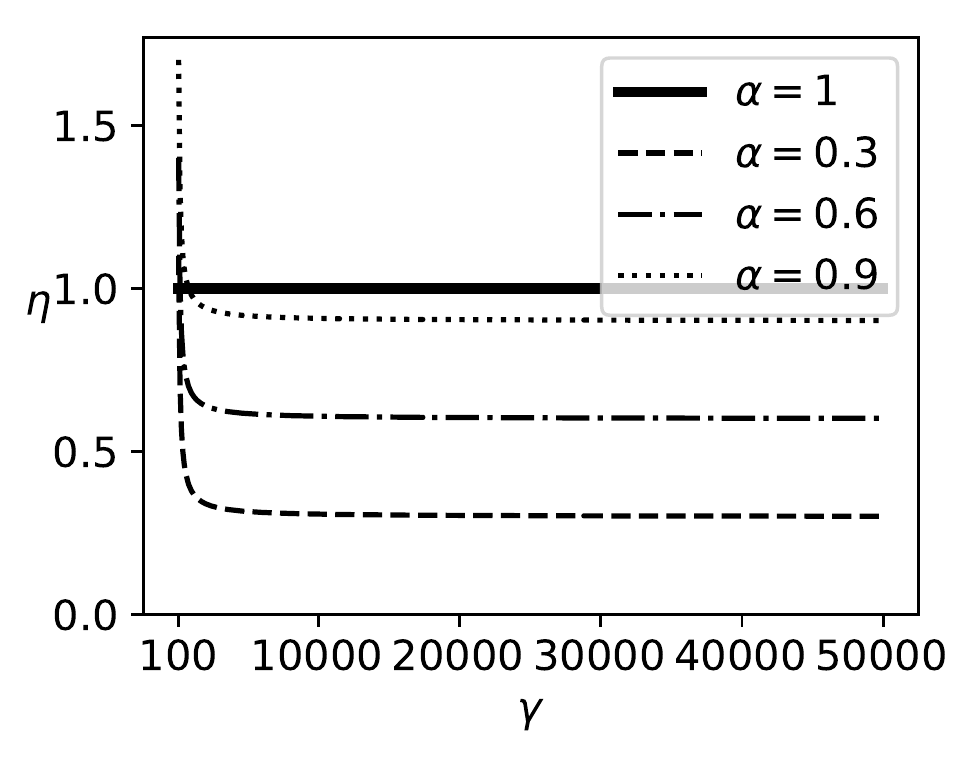}
				\subcaption{Efficiency graph using $\eta = \alpha  + \rho(|\mathit{SB}|+ |\mathit{NSB}|)/\gamma$.}
				\label{fig:non-index-eta-graph-using-equation}
			\end{minipage}
			\begin{minipage}[b]{.50\linewidth}
				\centering
			\includegraphics[scale=0.46]{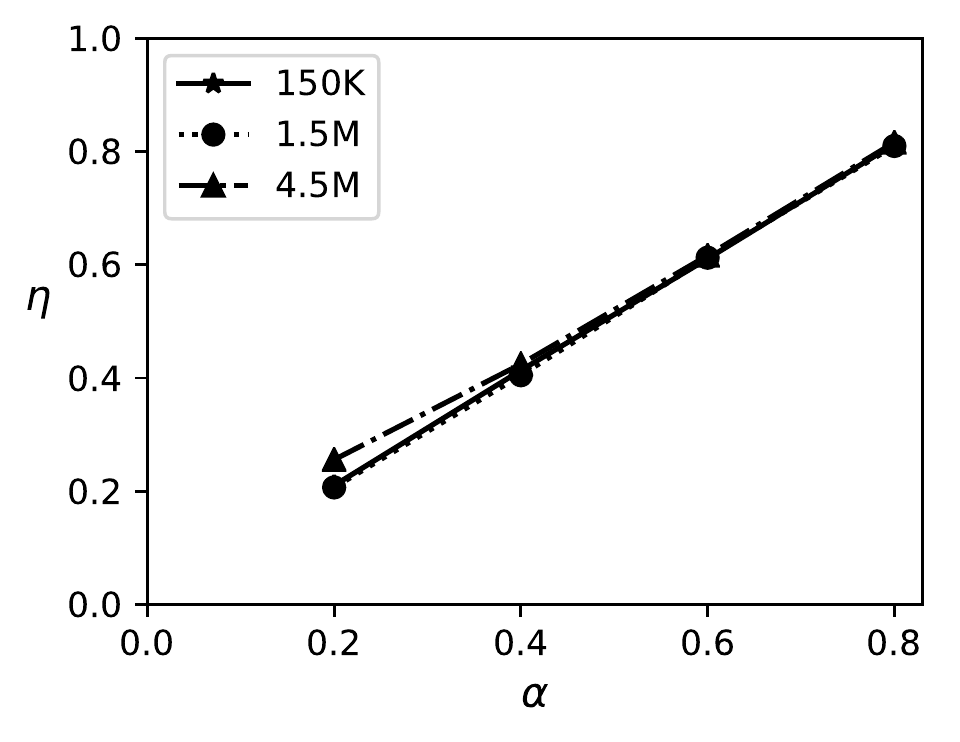}\B
			\subcaption{Dataset size.}
				\label{fig:data_set_size_increase}
			\end{minipage}
			\begin{minipage}[b]{.45\linewidth}
				\centering
				\includegraphics[scale=0.46]{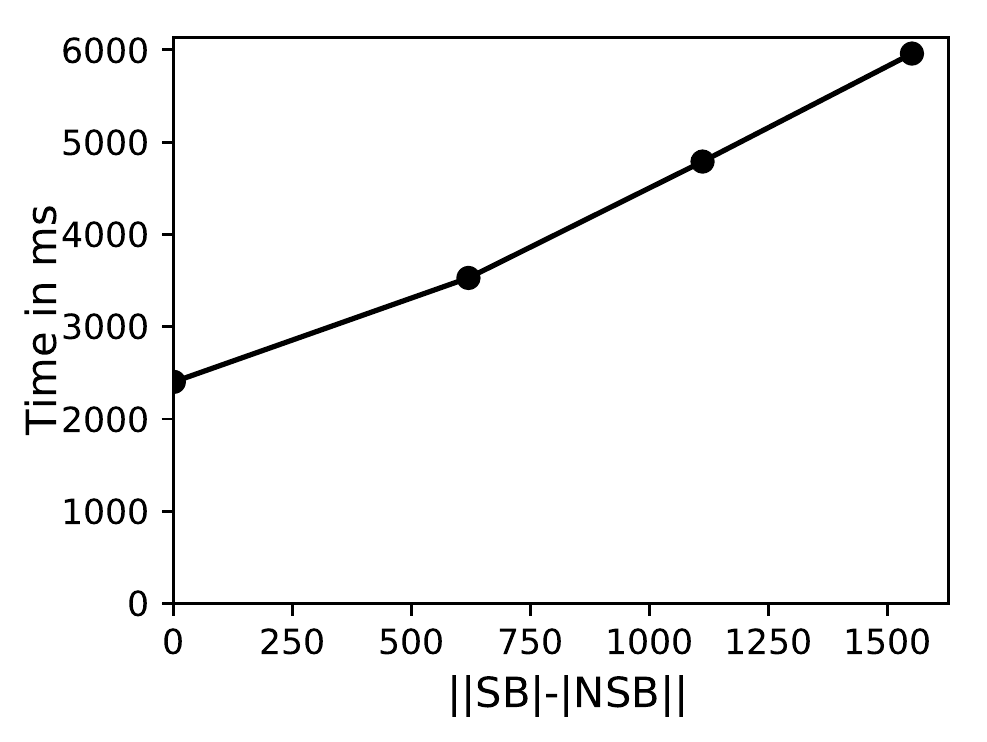}\B
				\subcaption{Bin Size.}
				\label{fig:vary_bin_size}
			\end{minipage}	
\end{center}
		\caption{Experiments.}
		\label{fig:Side_Effect_Test}
	\end{figure}

The equation above demonstrates that QB trades increased communication costs to reduce the amount of data that needs to be searched in encrypted form. Note that the reduction in encryption cost is proportional to $\alpha$ times the size of the database, while the increase in communication costs is proportional to $\sqrt{|D|}$, where $|D|$ is the number of distinct attribute values. This, coupled with the fact that $\gamma$ is much higher than 1 for encryption mechanisms that offer strong security guarantees, ensures that QB almost always outperforms the full encryption approaches. For instance, the cryptographic cost for search using secret-sharing is $\approx 10ms$ \cite{DBLP:journals/isci/EmekciMAA14}, while the cost of transmitting a single row ($\approx $ 200 bytes for TPCH Customer table) is $\approx 4$ $\mu$$s$ making the value of $\gamma \approx 25000$. Thus, QB, based on the model, should outperform the fully encrypted solution for almost any value of $\alpha$, under ideal situations where our assumption of uniformity holds. Figure~\ref{fig:non-index-eta-graph-using-equation} plots a graph of $\eta$ as a function of $\gamma$, for varying sensitivity and $\rho = 10\%$.


\subsection{Experimental Validation}
\noindent We determined $\eta$ values for two commercial databases that support non-deterministic encryption. We refer to them as systems A and B, respectively, to hide their identities.

\noindent\textbf{Search Techniques.} To support
encrypted search on both systems, since they do not provide a searching facility on the non-deterministically encrypted data, we implemented the following technique: retrieves the searching attribute of a sensitive relation at the DB owner side, decrypts the attributes, and searches for records that match $|\mathit{SB}|$. It then retrieves full tuples corresponding to $|\mathit{SB}|$ predicates' addresses. For comparing against cryptographic searches at the cloud-side, we used SGX-based Opaque~\cite{opaque} and the multi-party computations (MPC) based Jana~\cite{jana} for evaluating QB's effectiveness.

\noindent\textbf{Experimental setup.} We used a virtual machine of 2.6 GHz, 4 core processor, 16 GB RAM, 1TB disk, and average 30Mbps download speed. We used TPCH benchmark to generate the dataset. 
The DB owner stores sensitive and non-sensitive bins, whose size was propositional to the domain size of the searchable attributes and independent of the database size. For TPC-H LINEITEM table, metadata for attributes L\_PARTKEY and L\_SUPPKEY were 13.6MB and 0.65MB, respectively. 

\begin{table}[t]
  \centering
    \begin{tabular}{|l|l|l|l|l|l|l|}
    \hline
    Technique & 1\% & 5\% &     20\% & 40\% & 60\% \\ \hline
    SGX-based Opaque~\cite{opaque} & 11 & 15 &     26 & 42 & 59 \\ \hline
    MPC-based Jana~\cite{jana} & 22 & 80 &         270 & 505 & 749  \\ \hline
    \end{tabular}
    \caption{Time (in seconds) when mixing QB with Opaque and Jana at different levels of sensitivity.}
    \label{tab:opaque mpc}
\end{table}

\noindent\textbf{Exp 1: Robustness of QB.} To explore the effectiveness of QB under different DB sizes, we tested QB for 3 DB sizes: 150K, 1.5M, and 4.5M tuples using No-Ind(A) and No-Ind(B) as underlying cryptographic mechanisms. Figure~\ref{fig:data_set_size_increase} plots $\eta$ values for the three sizes for No-Ind(A) while varying $\alpha$. The figure shows that $\eta < 1$, irrespective of the DB sizes, confirming that QB scales to larger DB sizes (results over No-Ind(B) are similar). Table~\ref{tab:opaque mpc} shows the time taken when using QB with Opaque and Jana at different levels of sensitivity. Without using QB for answering a simple selection query, Opaque~\cite{opaque} took 89 seconds on a dataset of size 700MB (6M tuples) and Jana~\cite{jana} took 1051 seconds on a dataset of size 116MB (1M tuples). Note that the time to execute the same query on cleartext data of size 700MB took only 0.0002 seconds. QB improves not only the performance of Opaque and Jana, but also makes them to work securely on partitioned data and resilient to output-size attack. The performance of QB will be even higher when one uses more secure cryptographic techniques that are resilient to output-size attacks, since these techniques will consume significant time for answering a query.

\noindent\textbf{Exp 2: Effect of bin size.} Figure~\ref{fig:vary_bin_size} plots an average time for a selection query using QB with a different bin size, which is in turn governed by the values of $|\mathit{SB}|$ and $|\mathit{NSB}|$, respectively. We plot the effect of $||\mathit{SB}|-|\mathit{NSB}||$ on retrieval time and find that the minimum time is achieved when $|\mathit{SB}|=|\mathit{NSB}|$. Thus, the optimal choice is $|\mathit{SB}|=|\mathit{SB}|=\sqrt{|\mathit{NS}|}$. 

\section{Desiderata}
\label{sec:Desiderata}

Below we focus on an aspect of QB, which is to a degree surprising. While QB is designed for scaling cryptographic techniques, it has a side-effect of improving security properties of an underlying cryptographic technique. In particular, a cryptographic technique that is prone to output-size, frequency-count, and workload-skew attacks becomes secure against these attacks when mixed with QB. Thus, QB offers a higher level of security, in addition to saving the cryptographic search on non-sensitive data.

\noindent
\textbf{Enhancing security-levels of indexable techniques.} We show how QB can be integrated with an indexable cryptographic technique, namely Arx~\cite{arx-popa-2017} that uses a non-deterministic encryption mechanism. In Arx, the DB owner stores each domain value $v$ and the frequency of $v$ in the database. The technique encrypts the $i^{\mathit{th}}$ occurrence of $v$ as a concatenated string $\langle v,i\rangle$ thereby ensuring that no two occurrences of $v$ result in an identical ciphertext. Such a ciphertext representation can then be indexed on the cloud-side. During retrieval, the user keeps track of the histogram of occurrences for each value and generates appropriate ciphertexts that can be used to query the index on the cloud. It is not difficult to see that Arx, by itself, is susceptible to the size, frequency-count, workload-skew, and access-pattern attacks. The query processing using Arx as efficient as cleartext version due to using an index, \textit{e}.\textit{g}., $\beta$ values for Arx are $1.4$ on system A and $2.5$ on system B.

The use of QB with Arx makes it secure against output-size, frequency-count, and workload-skew attacks. Of course, QB takes more time as compared to Arx, since the time of $|\mathit{SB}|$ searches cannot be absorbed in a single index scan unless all $|\mathit{SB}|$ values lie in a single node of the index. In the worst case, we traverse the index at most $|\mathit{SB}|$ times, unlike Arx, which traverses the index only once for a single selection query. 
It, however, significantly enhances the security of Arx by preventing output size, frequency count, and workload-skew attacks. However, QB does not protect access-patterns being revealed which could be prevented using ORAM. Determining whether coupling ORAM with Arx mixed with QB or using a more secure cryptographic solution, \textit{e}.\textit{g}., secret-sharing, which uses a linear scan 
to prevent access-patterns, with QB, more efficient (while QB with both the solutions strengthen the underlying cryptographic technique) is an open question.

\section{Conclusion}
\label{sec:Conclusion}
We propose query binning (QB) technique that serves as a meta approach on top of existing cryptographic techniques to support secure selection queries when a relation is partitioned into cryptographically secure sensitive and clear-text non-sensitive sub-relations. Further, we develop a new notion of partitioned data security that restricts exposing sensitive information due to the joint processing of the sensitive and non-sensitive relations. Besides improving efficiency, while supporting partitioned security, interestingly, QB enhances the security of the underlying cryptographic technique by preventing size, frequency-count, and workload-skew attacks. Thus, combining QB with efficient but non-secure cloud-side indexable cryptographic approaches result in an efficient and significantly more secure search. Furthermore, existing access-pattern-hiding cryptographic techniques also benefit from the added security that QB offers. 


\end{document}